\documentclass[12pt]{article}
\usepackage{amsmath,amssymb,amsfonts,graphicx,bm}
\usepackage{pgfplots}
\usepackage{graphicx,bm}
\usepackage{geometry}
  
\renewcommand{\theequation}{\arabic{section}.\arabic{equation}}

    \newgeometry{vmargin={20mm,30mm}, hmargin={25mm,25mm}}  
    
    \usepackage[margin=1cm]{caption}

\newcommand{\R}{{\mathbb R}}

\newcommand{\x}{\mathbf{x}}

\newcommand{\y}{\mathbf{y}}
\newcommand{\X}{\bm{X}}

\newcommand{\E}{\mathbb E}

\newcommand{\calT}{\mathcal{T}}
\newcommand{\calA}{\mathcal{A}}

\newcommand{\z}{{\bf z}}
\newcommand{\e}{{\rm e}}

\renewcommand{\L}{{ \mathbb L}}

\usepackage{empheq}
\usepackage{framed}
\usepackage[most]{tcolorbox}
\usepackage{xcolor}
\usepackage{fancyhdr}
\newcommand{\pcb}[1]{\textcolor{black}{#1}}

\newcommand{\PP}{\widetilde{P}}
\renewcommand{\P}{\mathbb P}

\begin{document}

\title{A generalized Dean-Kawasaki equation for an interacting Brownian gas in a partially absorbing medium}
\author{ \em
Paul. C. Bressloff, \\ Department of Mathematics, Imperial College London, \\
London SW7 2AZ, UK}

\maketitle

\begin{abstract}
The Dean-Kawasaki (DK) equation is a stochastic partial differential equation (SPDE) for the global density $\rho$ of a gas of $N$ over-damped Brownian particles. In the thermodynamic limit $N\rightarrow \infty$ with weak pairwise interactions, the expectation $\E[\rho]$ converges in distribution to the solution of a McKean-Vlasov (MV) equation. In this paper we derive a generalized DK equation for an interacting Brownian gas in a partially absorbing one-dimensional medium. In the case of the half-line with a totally reflecting boundary at $x=0$, the generalized DK equation is an SPDE for the joint global density $\rho(x,\ell,t)=N^{-1}\sum_{j=1}^N\delta(x-X_j(t))\delta(\ell-L_j(t))$, where $X_j(t)$ and $L_j(t)$ denote the position and local time of the $j$th particle, respectively. Assuming the DK equation has a well-defined mean field limit, we derive the MV equation on the half-line with a reflecting boundary, and analyze stationary solutions for a Curie-Weiss (quadratic) interaction potential. We then use an encounter-based approach to develop the analogous theory for a partially absorbing boundary at $x=0$. Each particle is independently absorbed when its local time $L_j(t)$ exceeds a random threshold $\widehat{\ell}_j$ with probability distribution $\Psi(\ell)=\P[\widehat{\ell}_j>\ell]$. The joint global density is now summed over the set of particles that have not yet been absorbed, and expectations are taken with respect to the Gaussian noise and the random thresholds $\widehat{\ell}_j$. Extensions to finite intervals and partially absorbing traps are also considered.
 \end{abstract}

 \section{Introduction} The Dean-Kawasaki (DK) equation is a stochastic partial differential equation (SPDE) that describes hydrodynamic fluctuations in the global density $\rho(\x,t)=N^{-1}\sum_{j=1}^N\delta(\x-\X_j(t)) $ of $N$ over-damped Brownian particles (Brownian gas) with positions $\X_j(t)\in \R^d$ at time $t$ \cite{Dean96,Kawasaki98}. More specifically, suppose that the positions evolve according to the stochastic differential equation (SDE)
 \begin{equation}
 d\X_j(t)=-\frac{1}{N\gamma}\sum_{k=1}^N{\bm \nabla} K(\X_j(t)-\X_k(t)|)+\sqrt{2D}d{\bf W}_j(t),
 \end{equation}
 where $D$ is the diffusivity, $\gamma$ is a drag  
coefficient with $D\gamma=k_BT$, $K$ is a smooth pairwise potential, and ${\bf W}_j(t)$ is a vector of independent Wiener processes. The DK equation then takes the form \cite{Dean96}
 \begin{align}
  \frac{\partial \rho(\x,t)}{\partial t} 
&=\sqrt{\frac{2D}{N}}{\bm \nabla} \cdot \bigg [ \sqrt{ \rho(\x,t)} { \bm \eta}(\x,t)\bigg ]+D{\bm \nabla}^2 \rho(\x,t) \nonumber \\
 & \quad  +\frac{1}{\gamma} {\bm \nabla}\cdot \bigg (  \rho(\x,t) \int_{\R^d}    \rho(\y,t){\bm \nabla}K(\x-\y)d\y \bigg ) ,
\label{DKN}
\end{align}
where ${\bm \eta}(\x,t)$ is a vector of independent spatiotemporal white noise processes. 
Formally speaking, equation (\ref{DKN}) is an exact equation for the global density in the distributional sense. Although the solution of the DK equation is highly singular, it provides a basis for accurate and efficient numerical
simulations of the density fluctuations of independent diffusing particles \cite{Cornalba23}. The exact density equation has also been used to construct a statistical field theory of a non-interacting Brownian gas \cite{Velenich08}. If particle-particle interactions are included, then averaging the DK equation with respect to the Gaussian noise processes results in a moment closure problem for the one-particle density $\E[\rho]$. One approximation scheme for achieving moment closure, which is used extensively in non-equilibrium statistical physics, is dynamical density functional theory (DDFT) \cite{Marconi99,Evans04,Archer04,Witt21}. A crucial assumption of DDFT is that the relaxation of the system is sufficiently slow such that the pair correlation can be equated with that of a corresponding equilibrium system at each point in time. An alternative approach is to use mean field theory. There is an extensive mathematical literature on the rigorous stochastic analysis of the mean field limit $N\rightarrow \infty$ for weak pairwise interactions, see for example Refs. \cite{Oelsch84,Jabin17,Chaintron22a,Chaintron22b}. 
In particular, if the initial positions of the $N$ particles are independent and identically distributed, i.e. the joint probability density at $t=0$ takes the product form
$p(\x_1,\ldots,\x_{N},0)=\prod_{j=1}^{N}\phi_0(\x_j)$, then it can be proven that $\E[\rho]$ converges in distribution to the solution of the McKean-Vlasov (MV) equation \cite{McKean66}
\begin{eqnarray}
 \frac{\partial \phi(\x,t)}{\partial t} 
=D{\bm \nabla}^2 \phi(\x,t)  +\frac{1}{\gamma}{\bm \nabla}\cdot  \bigg ( \phi(\x,t) \int_{\R^d} \phi(\y,t) {\bm \nabla}K(\x-\y)d\y\bigg ),
\label{MV0}\end{eqnarray}
with $\phi(\x,0)=\phi_0(\x)$. Equation (\ref{MV0}) has an alternative interpretation as the nonlinear FP equation for the so-called nonlinear McKean SDE
\begin{equation}
d\X=-\frac{1}{\gamma} \left [\int_{\R^d}{\bm \nabla }K(\X(t)-\y|)\rho(\y,t)d\y\right ]dt+\sqrt{2D}d{\bf W}(t).
\end{equation}
The interacting Brownian gas is said to satisfy the propagation of chaos property. 
The MV equation is known to have a rich mathematical structure, which includes the existence of multiple stationary solutions and associated phase transitions \cite{Tamura84}. This has been explored in various configurations, including double-well confinement and Curie-Weiss interactions on $\R$ \cite{Desai78,Dawson83,Pavliotis19}, and interacting particles on a torus \cite{Chayes10,Carrillo20}.  

Most studies of interacting Brownian gases ignore the effects of boundaries, with a few notable exceptions that consider the mean field limit in the presence of reflecting  boundaries \cite{Sznitman84,Coghi22}. There have also been a few studies of absorbing boundaries within the contexts of mathematical finance \cite{Hambly17} and mean field games \cite{Campi18,Caines20}.
In this paper we derive a generalized DK equation for a weakly interacting Brownian gas in a partially absorbing one-dimensional medium. We begin by considering diffusion on the half-line with a reflecting boundary at $x=0$ (section 2). The generalized DK equation takes the form of an SPDE for the joint global density 
\begin{equation}
\rho(x,\ell,t)=\frac{1}{N}\sum_{j=1}^N\delta(x-X_j(t))\delta(\ell-L_j(t)), 
\end{equation}
given the positions $X_j(t)$ and local times $L_j(t)$ of the particles, $j=1,\ldots,N$. The local time is a Brownian functional that characterizes the amount of time that a Brownian particle spends in the neighborhood of a totally reflecting boundary \cite{Levy39,Ito63,Ito65,McKean75,Majumdar05}. Heuristically speaking, the differential of the local time generates an impulsive kick whenever the particle encounters the boundary, whose inclusion leads to the stochastic Skorokhod equation for reflected Brownian motion \cite{Freidlin85}. We show that the DK equation in the bulk domain $(0,\infty)$ and the boundary condition at $x=0$ include a nonlocal term that depends on the reduced field $\overline{\rho}(x,t)=\int_0^{\infty}\rho(x,\ell,t)d\ell$ and a multiplicative noise term that depends on $\sqrt{\rho(x,\ell,t)}$. Assuming the SPDE for $\rho$ has a well-defined mean field limit, we derive a nonlinear Fokker-Planck equation for $\E[\rho]$, which is then used to derive a corresponding MV equation for $\E[\overline{\rho}]$. We thus recover the MV equation for reflected diffusions previously obtained using methods from stochastic analysis \cite{Sznitman84,Coghi22}. The straightforward extension to a Brownian gas on a finite interval is also described.
In section 3 we consider the stationary solutions of the MV equation in the case of a Curie-Weiss (quadratic) interaction potential for both the semi-infinite and finite intervals. In the latter case, we explore how the existence of phase transitions depends on the size of the domain. 

In section 4 we combine the generalized DK equation with an encounter-based model of a partially absorbing boundary at $x=0$ \cite{Grebenkov20,Grebenkov22,Bressloff22,Bressloff22a}. Each particle is independently absorbed when its local time $L_j(t)$ exceeds a random threshold $\widehat{\ell}_j$ with probability distribution $\Psi(\ell)=\P[\widehat{\ell}_j>\ell]$. The corresponding global joint density $\mu$ only sums over the set of particles that haven't yet been absorbed, that is, 
\begin{equation}
\mu(x,\ell,\widehat{\bm \ell},t)=\frac{1}{N}\sum_{j=1}^N\delta(x-X_j(t))\delta(\ell-L_j(t)){\bf 1}_{L_j(t)<\widehat{\ell}_j}. 
\end{equation}
We derive the generalized DK equation for $\mu$ and then use a mean field ansatz to obtain a MV equation for $\E[\overline{\mu}]$, $\overline{\mu}(x,\widehat{\bm \ell},t)=\int_0^{\infty} \mu(x,\ell,\widehat{\bm \ell},t)d\ell$, where expectation is taken with respect to the Gaussian noise processes and the random local time thresholds. The MV equation depends on the choice of distribution $\Psi$ such that $\E[\overline{\mu}]=\int_0^{\infty} \Psi(\ell)\phi(x,\ell,t)d\ell$ for some unknown function $\phi(x,\ell,t)$. A complicating factor is that the boundary condition at $x=0$ equates the particle flux with the rate of absorption, which is given by a term proportional to $\int_0^{\infty} \psi(\ell)\phi(0,\ell,t)$, where $\psi(\ell)=-\Psi'(\ell)$. Hence, for a general threshold distribution $\Psi$, the MV equation is not a closed equation for
$\E[\overline{\mu}]$. One important exception is the exponential distribution, $\Psi(\ell)=\exp(-\kappa_0\ell/D)$, for which the boundary condition is of Robin type and $\E[\overline{\mu}]$ is equivalent to the Laplace transform $\widetilde{\phi}(x,z,t)$,  with respect to $\ell$ and $z=\kappa_0/D$. Hence, assuming a solution of the nonlinear Robin boundary value problem (BVP) exists, the corresponding function $\phi(x,\ell,t)$ can be determined by inverting the Laplace transform, which then determines $\E[\overline{\mu}]$ for a general $\Psi$ by integration. We illustrate the theory by considering the effective rate of particle loss in the weak absorption limit. Finally, in section 5 we describe various possible extensions of the theory, including an interacting Brownian gas in $\R$ with a finite interval acting as a partially absorbing trap. Absorption is now conditioned on the occupation time (time spent within the trapping region) crossing a random threshold \cite{Bressloff22,Bressloff22a}.

    \setcounter{equation}{0}

\section{Generalized DK equation for a totally reflecting boundary} 

In this section we derive the generalized DK equation for a Brownian gas on $[0,\infty)$ with a totally reflecting boundary at $x=0$. We begin by considering a single Brownian particle.

\subsection{Single Brownian particle}

Consider a single Brownian particle restricted to the half-line $[0,\infty)$ with a reflecting boundary at $x=0$. Let $L(t)$ be the boundary local time, which is a Brownian functional of the form
\begin{equation}
L(t)=\lim_{\epsilon\rightarrow 0^+}\frac{D}{\epsilon}\int_0^t{\bm 1}_{(0,\epsilon)}(X(s)) ds,
\end{equation}
where ${\bf 1}$ is the indicator function. (The factor of $D$ means that $L(t) $ has units of length.) It can be proven that $L(t)$ exists and is a nondecreasing, continuous function of $t$ \cite{Ito63,Ito65}. 
  The SDE for $X(t)\in [0,\infty)$ is given by the so-called Skorokhod equation for reflecting Brownian motion,     \begin{align}
        \label{sde-outside}
        dX(t)= \sqrt{2D}dW(t)+ dL (t).    \end{align}
 Formally speaking, $dL(t)=D\delta(X(t))dt$ so that each time the particle hits a boundary it is given an impulsive kick back into the domain in a direction perpendicular to the boundary. Consider 
the joint probability density or local time propagator for the pair $(X(t),L(t))$:
\[P(x,\ell,t)dx\, d\ell :=\P[x\leq X(t) <x+dx, \ell \leq  L(t)<\ell+d\ell ].\]
Since the local time only changes at the membrane boundary $x=0$, the evolution equation within the bulk of the domain is simply 
\begin{subequations}
\label{JPCK1}
 \begin{align} 
   \frac{\partial P}{\partial t}  = D\frac{\partial^2 P}{\partial x^2},\ x>0, \ \ell \geq 0, \ t>0.
     \end{align}
     However, the boundary condition at $x=0$ becomes \cite{Grebenkov06}.
     \begin{equation}
\left .\frac{\partial P(x,\ell,t)}{\partial x}\right |_{x=0}=P(0,0,t)\delta(\ell)+\frac{\partial P(0,\ell,t)}{\partial \ell}.
\end{equation}
\end{subequations}
Integrating equations (\ref{JPCK1}) with respect to $\ell$ then recovers the standard diffusion equation for the marginal density $p(x,t)=\int_0^{\infty}P(x,\ell,t)d\ell$ with a Neumann boundary condition at $x=0$:
\begin{align} 
 \frac{\partial p(x,t)}{\partial t} 
=D\frac{\partial^2 p(x,t)}{\partial x^2} ,\
\left . D\frac{\partial p(x,t) }{\partial x}\right |_{x=0}=0  .
\end{align}

\subsection{Non-interacting Brownian gas}

Suppose that there are now $N$ identical, non-interacting Brownian particles on the half-line.  Each particle is subject to a totally reflecting boundary at $x=0$ so that it accumulates its own local time $L_j(t)$, $j=1,\ldots,N$. The position $X_j(t)$ of the $j$th particle evolves according to the SDE
\begin{equation}
\label{multiR}
dX_j(t)=\sqrt{2D}d{W}_j(t)+dL_j(t),\quad dL_j(t)=D\delta(X_j(t))dt,
 \end{equation}
 with $W_j(t)$, $j=1,\ldots,N$, a set of independent Wiener processes. A compact description of the dynamics can be obtained by considering a ``hydrodynamic'' formulation of equation (\ref{multiR}), which involves the (normalized)
 global density
\begin{equation}
\label{global}
\rho(x,\ell,t)=\frac{1}{N}\sum_{j=1}^{N}\rho_j(x,\ell,t),\quad \rho_j(x,\ell,t)=\delta(X_j(t)-x)\delta(L_j(t)-\ell).
\end{equation}
The local time propagator of the $j$th particle can be expressed as
\begin{align}
\label{LTP}
P_j(x,\ell,t)=\bigg \langle \delta(X_j(t)-x)\delta(L_j(t)-\ell)\bigg \rangle,
\end{align}
where expectation is taken with respect to the white noise process.\footnote{Throughout the paper we use $\langle \cdot \rangle$ to denote expectation with respect to the Gaussian noise processes. In the analysis of partially absorbing boundaries in section 4, we use $\E[\cdot]$ to represent expectation with respect to a set of random local time thresholds.} 

We construct an SPDE for the global density $\rho$ by generalizing the derivation of the Dean-Kawasaki equation for a Brownian gas in $\R$ \cite{Dean96,Kawasaki98}. Consider an arbitrary smooth test function $f(x,\ell)$ with $\partial_xf(x,\ell)=0$ at $x=0$.
 Using Ito's lemma to Taylor expand $f(X_i(t+dt),L_i(t+dt))$ about $(X_i(t),L_i(t))$ and setting 
 \begin{equation}
f(X_i(t),L_i(t))=\int_{0}^{\infty}dx\int_0^{\infty}d\ell\,  \rho_i(x,\ell,t)f(x,\ell),
\end{equation}
we find that
\begin{align}
\label{dfl}
 \frac{df(X_i,L_i)}{dt}&=\int_{0}^{\infty}dx\int_{0}^{\infty}d\ell\, f(x,\ell)\frac{\partial \rho_i(x,\ell,t)}{\partial t} \\
 &=\int_{0}^{\infty}dx\int_{0}^{\infty}d\ell\, \rho_i(x,\ell,t)\bigg [\sqrt{2D}\partial_x f(x,\ell)  \xi_i(t)+D\partial_{xx} f(x,\ell)+D\partial_{\ell}f(x,\ell)\delta(x)\bigg ]. \nonumber
\end{align}
We have formally set $dW_i(t)=\xi_i(t)dt$ where $ \xi_i $ is a $d$-dimensional white noise term such that
\begin{equation}
\langle { \xi}_i(t)\rangle =0,\quad \langle {\xi}_i(t){\xi}_j(t')\rangle =\delta(t-t')\delta_{i,j}.
\end{equation}
Integrating by parts the various terms on the second line of equation (\ref{dfl}) gives
\begin{align}
&\int_{0}^{\infty}dx\int_{0}^{\infty}d\ell\, f(x,\ell)\frac{\partial \rho_i(x,\ell,t)}{\partial t}\nonumber \\
 &=\int_{0}^{\infty}dx\int_{0}^{\infty}d\ell\,\bigg [ f(x,\ell)\left (-\sqrt{2D}\partial_x\rho_i(x,\ell,t)   \xi_i(t)+D\partial_{xx} \rho_i(x,\ell,t)\right )\bigg ]\nonumber \\
&\quad -\int_{0}^{\infty} \rho_i(0,\ell,t)\left (\sqrt{2D} f(0,\ell)  \xi_i(t)+D\partial_{x} f(0,\ell)\right )d\ell  \nonumber \\
&\quad -D\int_0^{\infty} f(0,\ell) \left [\partial_{\ell}\rho_i(x,\ell,t)-\partial_x\rho_i(0,\ell,t)\right ]d\ell -D \rho_i(0,0,t) f(0,0).
\label{RHS}
 \end{align}
Imposing the boundary condition $\partial_xf(0,\ell)=0$ and using the fact that $f(x,\ell)$ is otherwise arbitrary, we obtain the following equation for $\rho_i$:
\begin{subequations}
\label{rhoi}
\begin{align}
\frac{\partial \rho_i(x,\ell,t)}{\partial t} 
&=-\sqrt{2D}\frac{\partial \rho_i(x,\ell,t)}{\partial x}   \xi_i(t)+D\frac{\partial^2\rho_i(x,\ell,t)}{\partial x^2}+\delta(x){\mathcal J}(\ell,t),
\end{align}
with
\begin{align}
{\mathcal J}(\ell,t)\equiv D\frac{\partial \rho_i(0,\ell,t)}{\partial x}-D\frac{\partial \rho_i(0,\ell,t)}{\partial \ell} -\sqrt{2D} \rho_i(0,\ell,t)  \xi_i(t)-D \rho_i(0,0,t)  \delta(\ell)  .
\end{align}
\end{subequations}
Thus $\rho_i(x,\ell,t)$ has to satisfy the boundary condition ${\mathcal J}(\ell,t)=0$ at $x=0$.
The latter ensures conservation of particle number.

Summing equations (\ref{rhoi}) over the particle index $i$ and using the definition of the global density then gives
\begin{subequations}
\label{toto}
\begin{align}
\frac{\partial \rho(x,\ell,t)}{\partial t} 
&=-\frac{\sqrt{2D}}{N}\sum_{i=1}^{N}\frac{\partial \rho_i(x,\ell,t)}{\partial x}   \xi_i(t)+D\frac{\partial^2 \rho(x,\ell,t)}{\partial x^2},\\
D\frac{\partial \rho(0,\ell,t)}{\partial x} &=D\frac{\partial \rho(0,\ell,t)}{\partial  \ell} +\frac{\sqrt{2D}}{N}\sum_{i=1}^{N} \rho_i(0,\ell,t)  \xi_i(t)+D\rho(0,0,t)  \delta(\ell)  .
\end{align}
\end{subequations}
Following along analogous lines to Ref. \cite{Dean96}, we introduce the space-dependent Gaussian noise
\begin{equation}
\xi(x,\ell,t)=-\frac{1}{N}\sum_{i=1}^{N} \bigg [ \partial_x\rho_i(x,\ell,t)   \xi_i(t)\bigg ],
\end{equation}
with zero mean and the correlation function
\begin{equation}
\langle 
\xi (x,\ell,t)\xi (y,\ell',t')\rangle =\frac{1}{N^2} \delta(t-t')\sum_{i=1}^{N} \partial_x\partial_y\bigg (\rho_i(x,\ell,t) \rho_i(y,\ell',t) \bigg ).
\end{equation}
Since $\rho_i(x,\ell,t) \rho_i(y,\ell',t) =\delta(x-y)\delta(\ell-\ell')\rho_i(x,\ell,t)$, it follows that
\begin{equation}
\langle 
\xi (x,\ell,t)\xi (y,\ell',t')\rangle =\frac{1}{N} \delta(t-t')\delta(\ell-\ell'  ) \partial_x\partial_y \bigg (\delta(x-y) \rho(x,\ell,t)  \bigg ).
\end{equation}
Finally, we introduce the global density-dependent noise field
\begin{equation}
\widehat{\xi }(x,\ell,t)=\frac{1}{\sqrt{N}}\frac{\partial}{\partial x}\bigg (\eta(x,\ell,t)\sqrt{\rho(x,\ell,t)}\bigg ),
\end{equation}
where ${\eta}(x,\ell,t)$ is a spatiotemporal white noise term:
\begin{equation}
\langle  \eta(x,\ell,t)\eta(y,\ell',t')\rangle =\delta(t-t')\delta(x-y)\delta(\ell-\ell').
\end{equation}
It can be checked that the Gaussian noises ${ \xi}$ and $\widehat{ \xi}$ have the same correlation functions and are thus statistically identical. Hence, we obtain a closed SPDE for the global density:
\begin{subequations}
\label{rhoc2}
\begin{align} 
 \frac{\partial \rho(x,\ell,t)}{\partial t} 
&=\sqrt{\frac{2D}{N}}\frac{\partial \sqrt{\rho(x,\ell,t)} \eta(x,\ell,t)}{\partial x}+D\frac{\partial^2 \rho(x,\ell,t)}{\partial x^2} ,\\
D\frac{\partial \rho(0,\ell,t) }{\partial x}&=D\frac{\partial \rho(0,\ell,t)}{\partial \ell} -\sqrt{\frac{2D}{N}} \sqrt{\rho(0,\ell,t)} \eta(0,\ell,t) +D\rho(0,0,t)  \delta(\ell) .
\end{align}
\end{subequations}
Note that averaging with respect to the white noise and setting $\phi(x,\ell,t)=\langle \rho(x,\ell,t)\rangle $ recovers the evolution equation for the local time propagator, see (\ref{JPCK1}). (However, the initial conditions differ as $\phi$ arises from a multi-particle model.) Equation (\ref{rhoc2}) is the generalized DK equation for the global density $\rho(x,\ell,t)$ in the absence of particle interactions.

Integrating both sides of equation (\ref{rhoc2}) with respect to $\ell$ yields a corresponding DK equation for the marginal density $\overline{\rho}(x,t)=\int_0^{\infty}\rho(x,\ell,t)d\ell$:
\begin{subequations}
\label{rhocPsi0}
\begin{align} 
 \frac{\partial \overline{\rho}(x,t)}{\partial t} 
&=\sqrt{\frac{2D}{N}}\frac{\partial}{\partial x} \int_0^{\infty} \sqrt{\rho(x,\ell,t)} \eta(x,\ell,t)d\ell +D\frac{\partial^2 \overline{\rho}(x,t)}{\partial x^2} ,\\
D\frac{\partial \overline{\rho}(0,t) }{\partial x}&=-\sqrt{\frac{2D}{N}}\int_0^{\infty}  \sqrt{\rho(0,\ell,t)} \eta(0,\ell,t)d\ell  .
\end{align}
\end{subequations}
Introduce the transformed Gaussian stochastic variable
\begin{equation}
\theta(x,t)=\int_0^{\infty}\sqrt{\rho(x,\ell,t)} \eta(x,\ell,t)d\ell .
\end{equation}
We see that $\langle \theta(x,t)\rangle =0$ and
\begin{align}
\langle \theta(x,t)\theta(x',t')\rangle&=\int_0^{\infty}d\ell \int_0^{\infty}d\ell'\,    \sqrt{\rho(x,\ell,t)\rho(x',\ell',t')} \langle \eta(x,\ell,t) \eta(x',\ell',t')\rangle \nonumber \\
&=\delta(t-t')\delta(x-x') \int_0^{\infty}    \rho(x,\ell,t)d\ell=\delta(t-t')\delta(x-x')\overline{\rho}(x,t).
\end{align}
We can thus rewrite equations (\ref{rhocPsi0}) as
\begin{subequations}
\label{rhoc1}
\begin{align} 
 \frac{\partial \overline{\rho}(x,t)}{\partial t} 
&=\sqrt{2D}\frac{\partial}{\partial x} \bigg [\sqrt{\overline{\rho}(x,t)} \eta(x,t)\bigg ]+D\frac{\partial^2 \overline{\rho}(x,t)}{\partial x^2} ,\\
D\frac{\partial\overline{\rho}(0,t) }{\partial x}&=-\sqrt{2D}\sqrt{\overline{\rho}(0,t)} \eta(0,t) ,
\end{align}
\end{subequations}
where $\eta(x,t)$ is a scalar spatiotemporal white noise process. Finally, averaging with respect to the Gaussian noise results in the diffusion equation for $\overline{\phi}(x,t)=\langle\overline{\rho}(x,t) \rangle$ with a totally reflecting boundary at $x=0$.

The above derivations can also be applied to nonlinear functions of the density. For the sake of illustration, consider the equal-time correlation function
\begin{equation}
c(x,y,\ell,\ell',t)=\langle C(x,y,\ell,\ell',t)\rangle \equiv\langle \rho(x,\ell,t)\rho(y,\ell',t)\rangle.
\end{equation}
In appendix A we derive an SPDE for $C(x,y,\ell,\ell',t)$, which on averaging with respect to the spatiotemporal white noise yields a deterministic PDE for $c$, which takes the form
\begin{subequations}
\label{detCeq}
\begin{align}
\frac{\partial c(x,y,\ell,\ell',t)}{\partial t}&=
D\frac{\partial^2c(x,y,\ell,\ell',t)}{\partial x^2}+D\frac{\partial^2c(x,y,\ell,\ell',t)}{\partial y^2}\nonumber \\
&\quad +\frac{2D}{N}\delta(\ell-\ell')\frac{\partial^2}{\partial x\partial y}\delta(x-y)\phi(x,\ell,t),\quad x>0,\ y>0,
\end{align}
together with the boundary conditions
\begin{align}
 D\frac{\partial c(0,y,\ell,\ell',t)}{\partial x}&=D\frac{\partial c(0,y,\ell,\ell',t)}{\partial \ell} +D c(0,y,0,\ell',t)  \delta(\ell),\ y>0, \\
D\frac{\partial c(x,0,\ell,\ell',t)}{\partial y}&=D\frac{\partial c(x,0,\ell,\ell',t)}{\partial \ell'} +Dc(x,0,\ell,0,t)  \delta(\ell') ,\quad x>0  .
\end{align}
\end{subequations}
Similar to the analysis of the original DK equation \cite{Dean96}, the PDE for the correlation function couples to the average density $\phi=\langle \rho\rangle $.

\subsection{Interacting Brownian gas}

We now modify the SDE (\ref{multiR}) by introducing an external potential $V(x)$ and a pairwise interaction potential $K(x)$ such that
\begin{eqnarray}
 dX_j(t)=-\frac{1}{\gamma}\bigg [\partial_xV(X_j(t))+N^{-1}\sum_{k=1}^{N}\partial_x K(X_j(t)-X_k(t))\bigg ]dt+\sqrt{2D}d{W}_j(t)+dL_j(t).
\label{moo2}
 \end{eqnarray}
 The potentials contribute extra terms on the right-hand side of equation (\ref{dfl}) given by
\begin{align}
\calA_i&=-\frac{1}{\gamma }\int_{0}^{\infty}dx\int_{0}^{\infty}d\ell\, \rho_i(x,\ell,t) \left [\partial_xV(x)+\frac{1}{N}\sum_k \int_{0}^{\infty}dx' \delta(x'-X_k(t))\partial_xK(x-x')\right ] \nonumber \\
&\hspace{6cm} \times\partial_xf(x,\ell).
\end{align}
Integrating by parts with respect to $x$ and summing over $i$ yields
\begin{align}
\calA=\frac{1}{N}\sum_{i=1}^N \calA_i=\int_{0}^{\infty}dx \int_{0}^{\infty}d\ell\, \partial_x H_{\rho}(x,\ell,t)f(x,\ell)+ \int_{0}^{\infty}H_{\rho}(0,\ell,t)f(0,\ell)d\ell,
\end{align}
with
\begin{align}
H_{\rho}(x,\ell,t)&= \gamma^{-1} \rho(x,\ell,t) \bigg (\partial_xV(x)+\int_{0}^{\infty}\overline{\rho}(y,t)\partial_xK(x-y)dy \bigg ).\end{align}
The non-interacting terms are calculated along identical lines to the derivation of equations (\ref{rhoc2}), which leads to the following generalized DK equation for the interacting Brownian gas, which we write in the form of a conservation equation:
\begin{subequations}
\label{rhoc2int}
\begin{align} 
 \frac{\partial \rho(x,\ell,t)}{\partial t} &=-\frac{\partial  J(x,\ell,t)}{\partial x} ,\\
- J(0,\ell,t)&=D\frac{\partial \rho(0,\ell,t)}{\partial \ell}  +D\rho(0,0,t)  \delta(\ell),
 \end{align}
 with the probability flux
 \begin{align}
 J(x,\ell,t)&=-\sqrt{\frac{2D}{N}} \sqrt{\rho(x,\ell,t)} \eta(x,\ell,t) -D\frac{\partial \rho(x,\ell,t)}{\partial x} -H_{\rho}(x,\ell,t).
 \end{align}
\end{subequations}
Using similar arguments to the derivation of equations (\ref{rhoc1}), $\overline{\rho}$ evolves according to the equations
\begin{subequations}
\label{rhoc1b}
\begin{align} 
 \frac{\partial \overline{\rho}(x,t)}{\partial t} &=-\frac{\partial \overline{J}(x,t)}{\partial x},\quad  \overline{J}(0,t)=0,\\
\overline{J}(x,t)&=-\sqrt{\frac{2D}{N}}  \bigg [\sqrt{\overline{\rho}(x,t)} \eta(x,t)\bigg ]-D\frac{\partial \overline{\rho}(x,t)}{\partial x} -H_{\overline{\rho}}(x,t),
\end{align}
with
\begin{align}
H_{\overline{\rho}}(x,t)&=  \gamma^{-1}   \overline{\rho}(x,t) \bigg (\partial_xV(x)+\int_{0}^{\infty}\overline{\rho}(y,t)\partial_xK(x-y) dy \bigg ) . \end{align}
\end{subequations}
Equation (\ref{rhoc1b}) is precisely the DK equation one would expect to write down, given the original version defined on $\R$ \cite{Dean96} with the noise term included in the definition of the probability flux. However, in order to derive equation (\ref{rhoc1b}) from first principles, it is necessary to keep track of the local time of each particle.

Consistent with the classical DK equation (\ref{DKN}), averaging equation (\ref{rhoc2int}) with respect to the Gaussian noise processes leads to a PDE that couples the one-particle density $\phi(x,\ell,t)=\langle {\rho}(x,\ell,t)\rangle$ to the two-point correlation function $\langle {\rho}(x,\ell,t){\rho}(x',\ell',t)\rangle$ etc., resulting in a moment closure problem. The same issue applies to equations (\ref{rhoc2int}).
This raises the interesting problem of how to extend DDFT or mean field theory to handle moment closure in the case of the full global position and local time density $\rho(x,\ell,t)$. In this paper, we will assume that for sufficiently large $N$, the mean field approximation $\langle{\rho}(x,\ell,t) {\rho}(x',\ell',t)\rangle=\phi(x,\ell,t)\phi(x',\ell',t)$ holds. We thus obtain the following generalized MV equation for an interacting gas on the half-line with a totally reflecting boundary at $x=0$:
\begin{subequations}
\label{MVmft}
\begin{align}
 \frac{\partial \phi(x,\ell,t)}{\partial t} &=-D\frac{\partial J(x,\ell,t) }{\partial x},\\
 -J(0,\ell,t)&=D\frac{\partial \phi(0,\ell,t)}{\partial \ell}  +D\phi(0,0,t)  \delta(\ell),
\end{align}
with
\begin{align}
J(x,\ell,t)&=-D\frac{\partial \phi(x,\ell,t) }{\partial x}  -\frac{1}{\gamma} \bigg [ \phi(x,\ell,t) \bigg (\partial_xV(x)+\int_{0}^{\infty} \overline{\phi}(y,t) \partial_xK(x-y)dy\bigg )\bigg ],
\end{align}
\end{subequations}
Integrating both sides with respect to $\ell$ results in the reduced equation for $\overline{\phi}(x,t)=\int_0^{\infty}\phi(x,\ell,t)d\ell$:
\begin{subequations}
\label{MVbar2}
\begin{align}
 \frac{\partial \overline{\phi}(x,t)}{\partial t} &=-D\frac{\partial J(x,t) }{\partial x},\ x >0, \quad J(0,t)=0,\\
J(x,t)&=-D\frac{\partial \overline{\phi}(x,t) }{\partial x}  -\frac{1}{\gamma} \bigg [ \overline{\phi}(x,t) \bigg (\partial_xV(x)+\int_{0}^{\infty} \overline{\phi}(y,t) \partial_xK(x-y)dy\bigg )\bigg ].
\end{align}
\end{subequations}
This is equivalent to the MV equation derived previously by proving the propagation of chaos property in the thermodynamic limit \cite{Sznitman84,Coghi22}. 

\subsection{Brownian gas on a finite interval} Our derivation of the generalized DK equation can easily be extended to diffusion on the finite interval $[-R,R]$ with reflecting boundaries at both ends. The main modification is in the definition of the local time of the $j$th particle:
\begin{equation}
L_j(t)=\lim_{\epsilon\rightarrow 0^+}\frac{D}{\epsilon}\left [\int_0^tI_{(R-\epsilon,R)}(X_j(s))ds+\int_0^tI_{(-R,-R+\epsilon)}(X_j(s))ds\right ].
\label{local2}
\end{equation}
 The derivation of the corresponding DK equation for the global joint density $\rho(x,\ell,t)$ proceeds along similar lines to the half-line. In particular, equation (\ref{rhoc2int}) becomes
 \begin{subequations}
\label{rhofin}
\begin{align} 
 \frac{\partial \rho(x,\ell,t)}{\partial t} &=-\frac{\partial  J(x,\ell,t)}{\partial x} ,\quad x\in (-R,R),\\
-J(-R,\ell,t)&=D\frac{\partial \rho(-R,\ell,t)}{\partial \ell}  +D\rho(-R,0,t)  \delta(\ell)
,\\
J(R,\ell,t)&=D\frac{\partial \rho(R,\ell,t)}{\partial \ell}  +D\rho(R,0,t)  \delta(\ell),
 \end{align}
 \end{subequations}
 with the probability flux given by equation (\ref{rhoc2int}c) and
\begin{align}
H_{\rho}(x,\ell,t)&= \gamma^{-1} \rho(x,\ell,t) \bigg (\partial_xV(x)+\int_{-R}^{R}\overline{\rho}(y,t)\partial_xK(x-y)dy \bigg ).\end{align}
Averaging with respect to the Gaussians noise processes, imposing the mean field ansatz, and then averaging with respect to $\ell$ results in the following MV equation on the interval $[-R,R]$:
\begin{subequations}
\label{MVbar3}
\begin{align}
 \frac{\partial \overline{\phi}(x,t)}{\partial t} &=-D\frac{\partial \overline{J}(x,t) }{\partial x},\ x >0, \quad \overline{J}(-R,t)=0=\overline{J}(R,t),\\
\overline{J}(x,t)&=-D\frac{\partial \overline{\phi}(x,t) }{\partial x}  -\frac{1}{\gamma} \bigg [ \overline{\phi}(x,t) \bigg (\partial_xV(x)+\int_{-R}^{R} \overline{\phi}(y,t) \partial_xK(x-y)dy\bigg )\bigg ].
\end{align}
\end{subequations}

\section{Stationary states for the Curie-Weiss interaction potential}

For a finite system of interacting Brownian particles moving in a confining potential one finds that the associated linear FP equation has a unique stationary state given by the Boltzmann distribution. On the other hand, the MV equation is a nonlinear nonlocal FP equation that describes an interacting Brownian gas in the thermodynamic limit. Consequently, it can support the existence of multiple stationary solutions and their associated phase transitions \cite{Desai78,Dawson83,Tamura84,Chayes10,Pavliotis19,Carrillo20}. However, establishing the existence of a stationary solution of the MV equation is non-trivial, even in the absence of  boundaries. Here we explore this issue for the MV equation on the half-line and finite interval in the case of a Curie-Weiss (quadratic) interaction potential $K(x)=\lambda x^2/2$, $\lambda >0$.

\subsection{Stationary states on the half-line}  In the case of the half-line, the SDE (\ref{moo2}) reduces to the form
\begin{eqnarray}
 dX_j(t)=-\frac{1}{\gamma}\left [V'(X_j(t))+\frac{\lambda}{N}\sum_{k=1}^{N} [X_j(t)-X_k(t)]\right ]dt+\sqrt{2D}d{W}_j(t)+dL_j(t).
\label{mooDZ}
 \end{eqnarray}
 The interaction term can be rewritten as $-\lambda[X_j(t)-\overline{X}(t))$ where $\overline{X}=N^{-1}\sum_{k=1}^NX_k(t)$. It is an example of a cooperative coupling that tends to make the system relax towards the ``center of gravity'' of the multi-particle ensemble. If $V(x)$ is taken to be a multi-well potential then there is competition between the cooperative interactions and the tendency of particles to be distributed across the different potential wells according to the classical Boltzmann distribution.

The time-independent version of equation (\ref{MVbar2} is
\begin{equation}
\label{ssMVbar}
\frac{\partial {\overline{J}}(x)}{\partial x}=0,\ x >0,\quad J(0)=0,
\end{equation}
with
\begin{align}
\overline{J}(x):= -D\left [\frac{\partial \overline{\phi}(x) }{\partial x}  +\beta \overline{\phi}(x) \bigg ( V'(x)+\lambda \int_{0}^{\infty}(x-y) \overline{\phi}(y) dy\bigg )\right ].
\end{align}
We have used the Einstein relation $D\gamma =k_BT =\beta^{-1}$. Note that the integral term reduces to $\lambda (x-\langle y\rangle)$ with $\langle y\rangle =\int_0^{\infty}y\overline{\phi}(y)dy$. Suppose, for the moment, that $\langle y\rangle =a$ for some fixed $a$, which parameterizes the density $\overline{\phi}$. The totally reflecting boundary condition implies that $\overline{J}(x)=0$ for all $x\in [0,\infty)$ and, hence,
\begin{equation}
\label{ssbar}
\overline{\phi}=\overline{\phi}_a(x)=Z(a)^{-1}\exp\left (-\beta [V(x)+\lambda x^2/2-a\lambda x]\right ).
\end{equation}
The factor $Z(a)$ ensures the normalization $\int_0^{\infty} \overline{\phi}_a(x)dx=1$. 
The unknown parameter $a$ is then determined by imposing the self-consistency condition
\begin{equation}
\label{acon}
a=m(a)\equiv \int_0^{\infty} x\overline{\phi}_a(x)dx. 
\end{equation}
A necessary condition for the existence of a nontrivial solution $\phi_a(x)$ is that $V(x)+\lambda x^2/2\rightarrow 0$ as $x\rightarrow \infty$. The number of equilibrium solutions is then equal to the number of solutions of equation (\ref{acon}). Note that one major difference when diffusion is restricted to the half-line is that $a>0$ for any non-trivial solution $\overline{\phi}_a(x)$.

\begin{figure}[t!]
\centering
\includegraphics[width=16cm]{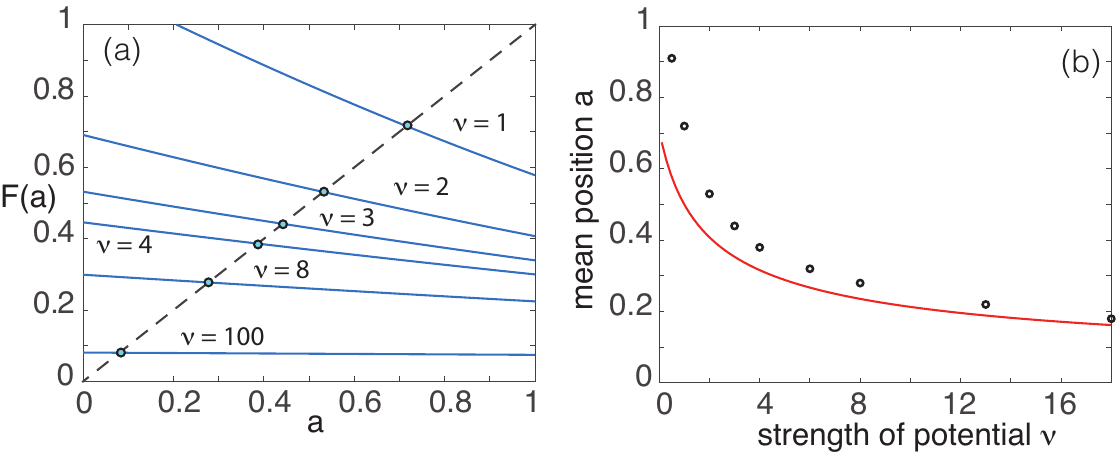} 
\caption{Brownian gas on the half-line. (a) Plot of function $F(a)$ defined in the the self-consistency condition (\ref{Fa}) for the mean position $a$ and various values of strength $\nu$ of the quadratic confining potential. The intercepts with the diagonal determine the unique solution $a$. Other parameters are $\lambda=\beta=1$. (b) Plot of intercepts as a function of $\nu$ for $\lambda=1$. The solid curve shows the mean position in the absence of coupling ($\lambda=0)$).}
\label{fig1}
\end{figure}

A common choice for $V$ in the case of a Brownian gas on $\R$ is the double-well potential $V(x)=x^4/4-x^2/2$. Although is not possible to analytically solve the corresponding equation $a=\int_{-\infty}^{\infty} x\overline{\phi}_a(x)dx$, one can prove that there exists a phase transition at a critical temperature $T_c$ such that $a=0$ for $T>T_c$ and $a=\pm a_0\neq 0$ for $T<T_c$ \cite{Desai78,Dawson83,Pavliotis19}. On the other hand, since the double well potential only has a single minimum in $[0,\infty)$ we expect there to exist at most one stationary solution for the reflected boundary problem. Therefore, we focus here on the existence of a unique stationary density for the simpler quadratic potential $V(x)=\nu x^2/2$, $\nu >0$. We then have
\begin{align}
Z(a)&=\int_0^{\infty} \e^{-\beta [(\nu+\lambda) x^2/2-a\lambda x]}dx\nonumber \\
&=\sqrt{\frac{\pi}{2\beta[ \nu +\lambda]}}\e^{\beta a^2\lambda^2/2[\nu +\lambda]}\mbox{erfc}(-a\lambda \sqrt{\beta/2[\nu+\lambda]}),
\label{Za}
\end{align}
and equation (\ref{acon}) becomes
\begin{align}
a &=Z(a)^{-1}\int_0^{\infty}x \e^{-\beta [(\nu +\lambda)x^2/2-a\lambda x]}dx=\frac{1}{\lambda\beta}\frac{\partial \log Z(a)}{\partial a}\nonumber \\
&= \frac{ a\lambda}{\nu+\lambda}+ \sqrt{\frac{2}{\pi \beta[ \nu+\lambda]}}\frac{\e^{-\beta a^2\lambda^2/2[\nu+\lambda] }}{\mbox{erfc}(-a\lambda \sqrt{\beta/2[\nu+\lambda]})}.
\end{align}
Rearranging this equation implies that $a$ is the implicit solution of
\begin{equation}
a=F(a):=\frac{\nu+\lambda}{\nu} \sqrt{\frac{2}{\pi \beta[ \nu+\lambda]}}\frac{\e^{-\beta a^2\lambda^2/2[\nu+\lambda] }}{\mbox{erfc}(-a\lambda \sqrt{\beta/2[\nu+\lambda]})}.
\label{Fa}
\end{equation}
In Fig. \ref{fig1}(a) we plot the function $F(a)$ for different values of $\nu$ and $\lambda=\beta=1$. This provides a graphical proof that there exists a unique stationary solution. The variation of the solution $a$ with $\nu$ is plotted in Fig. 1(b). We also compare with the mean position in the absence of interactions ($\lambda=0$). It can be seen that as $\nu\rightarrow \nu_c=0$, the effects of the cooperative interactions become more significant.

\begin{figure}[t!]
\centering
\includegraphics[width=16cm]{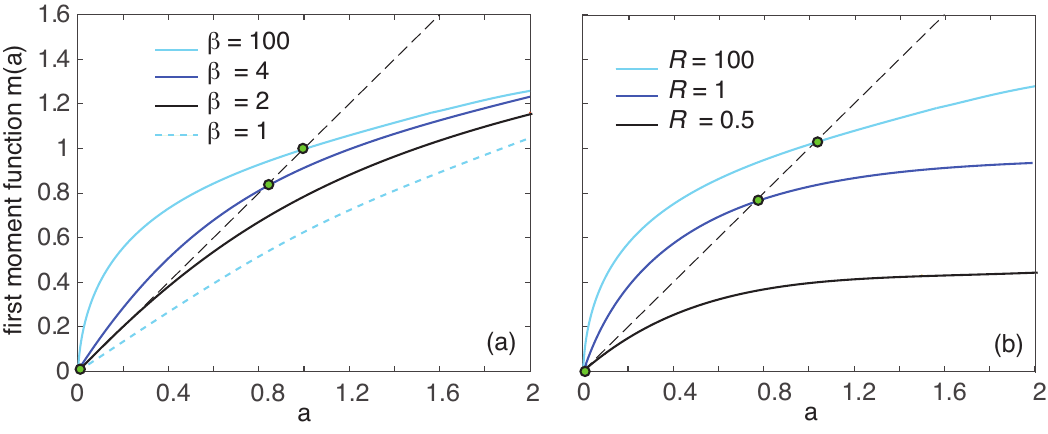} 
\caption{Brownian gas on $[-R,R]$. (a) Plot of first moment $m(a)$ as function of $a$ and various inverse temperatures $\beta$. The intercepts with the diagonal determine the positive solutions $a$. Other parameters are $\lambda=1$, $R=100$. (b) Corresponding plots for $\beta=10$ and various sizes $R$.}
\label{fig2}
\end{figure}

\subsection{Stationary states on $[-R,R]$}

The stationary solution (\ref{ssbar}) still holds in the finite interval except that $a$ is now a solution of the modified self-consistency condition
\begin{equation}
\label{acon2}
a=m(a)\equiv \int_{-R}^{R} x\overline{\phi}_a(x)dx.
\end{equation}
In the case of the quartic confining potential $V(x)=x^4/4-x^2/2$ we recover the results of Refs. \cite{Desai78,Dawson83} in the limit $R\rightarrow \infty$. That is, for sufficiently large $R$, there is a phase transition at a critical inverse temperature $\beta_c(R)$ between a single stationary state $a=0$ when $\beta <\beta_c$, and three stationary states $a=0,\pm a_0(\beta,L)$, $a_0>0$, when $\beta >\beta_c$) This is illustrated in Fig. \ref{fig2}(a) for $R=100$. We find numerically that $\beta_c \approx 2$ when $\lambda=1$, which is consistent with the critical point obtained in the limit $R\rightarrow \infty$ \cite{Desai78,Dawson83}. Our generalized mathematical framework allows us to explore how the phase transition depends on the size $R$ of the domain. As might be expected, for fixed $\beta>\beta_c^*$, where $\beta_c^*$ is the critical point in the limit $R\rightarrow \infty$, there exists a critical length $R_c(\beta)$ at which $a_0(\beta,R_c(\beta))=0$, see Fig. \ref{fig2}(b).

\section{Generalized DK equation for a partially absorbing boundary: an encounter-based model}

So far we have focused on totally reflecting boundary conditions, which requires keeping track of the local time of each particle. However, this information also allows us to incorporate a much more general class of boundary conditions via the encounter-based approach to diffusion-mediated surface absorption \cite{Grebenkov20,Grebenkov22,Bressloff22,Bressloff22a}. In this section we derive a generalized DK equation on the half-line with a partially absorbing boundary at $x=0$.

\subsection{Single Brownian particle}

At the single particle level, the encounter-based approach assumes that a diffusion process
is killed when the local time $L(t)$ at $x=0$ exceeds a randomly distributed threshold $\widehat{\ell}$. In other words, the particle is absorbed at $x=0$ at the stopping time
 \begin{equation}
\label{exp}
{\mathcal T}=\inf\{t>0:\ L(t)>\widehat{\ell}\},\quad \P[\widehat{\ell}>\ell]\equiv \Psi(\ell) .
\end{equation}
Since $L(t)$ is a nondecreasing process, the condition $t < {\mathcal T}$ is equivalent to the condition $L(t)<\widehat{\ell}$. Hence, the corresponding SDE is
\begin{eqnarray}
 dX(t)=[\sqrt{2D}dW(t)+dL(t)]{\bf 1}_{L(t)<\widehat{\ell}}.
 \end{eqnarray}
 In general, it is not possible to write down a closed differential equation for the corresponding marginal density  \begin{equation}
 \label{pmar}
p^{\Psi}(x,t)=\E\bigg [\bigg \langle \delta(x-X(t)\bigg \rangle\bigg ],
\end{equation}
where expectation is taken with respect to the Gaussian noise process and the random threshold $\widehat{\ell}$. Instead, one expresses $p^{\Psi}(x,t)$ in terms of the local time propagator $P(x,\ell,t)$, which is the solution to equations (\ref{JPCK1}). 
 \begin{align*}
p^{\Psi}(x,t)dx&=\P[x\leq X(t)<x+dx, \ L(t) < \widehat{\ell}]\\
&=\int_0^{\infty} d\ell \ \psi(\ell) \P[x\leq X(t)<x+dx, \ L(t)< \ell]\\
&=\int_0^{\infty} d\ell \psi(\ell)\int_0^{\ell} d\ell' [P(x,\ell',t)dx],
\end{align*}
where $\psi(\ell)=-\Psi'(\ell)$. Reversing the order of integration gives
\begin{align}
\label{pP}
p^{\Psi}(x,t)&=\int_0^{\infty}\Psi(\ell) P(x,\ell,t)d\ell .
\end{align}

Multiplying both sides of the propagator equations (\ref{JPCK1}a) and (\ref{JPCK1}b) by $\Psi(\ell)$ and integrating with respect to $\ell$ gives
\begin{subequations}
\label{pPsi}
\begin{align} 
 \frac{\partial p^{\Psi}(x,t)}{\partial t} 
&=D\frac{\partial^2 p^{\Psi}(x,t)}{\partial x^2} ,\\
\left . D\frac{\partial p^{\Psi}(x,t) }{\partial x}\right |_{x=0}&=D\int_0^{\infty} \psi(\ell) P(0,\ell,t)d\ell  .
\end{align}
\end{subequations}
For a general local time threshold distribution $\Psi$, we do not have a closed equation for the marginal density $p^{\Psi}(x,t)$. However, in the particular case of the exponential distribution $\Psi(\ell)=\e^{-\kappa_0\ell /D}$, $\psi(\ell)=\kappa_0 \Psi(\ell)/D$ and equations (\ref{pPsi}) reduce to the classical Robin BVP with reactivity $\kappa_0$:
 \begin{subequations}
\label{Robin}
\begin{align}
\frac{\partial p(x,t)}{\partial t}&=D\frac{\partial^2 p(x,t)}{\partial x^2}, \quad x> 0,\\
\left .D\frac{\partial p(0,t)}{\partial x}\right |_{x=0}&=\kappa_0 p(0,t).
\end{align}
\end{subequations}
(We have set $p^{\Psi}=p$ for $\Psi(\ell)=\e^{-\kappa_0\ell/D}$.) Within the context of the encounter-based formalism, the solution of the Robin BVP is equivalent to the Laplace transform of the propagator with respect to $\ell$:
\begin{equation}
p(x,t)=\int_0^{\infty} \e^{-z \ell} P(x,\ell,t)d\ell=\PP(x,z,t),\quad z=\kappa_0/D.
\end{equation}
Assuming that the Laplace transform $\PP(x,z,t),$ can be inverted with respect to $z$, the solution of equation (\ref{pPsi}) is obtained from equation (\ref{pP}): 
 \begin{equation}
  \label{oo}
  p^{\Psi}(x,t)=\int_0^{\infty} \Psi(\ell)P(x,\ell,t)d\ell = \int_0^{\infty} \Psi(\ell){\mathcal L}_{\ell}^{-1}\widetilde{P}(x,z,t)d\ell.
  \end{equation}
  One way to implement a non-exponential law is to consider an $\ell$-dependent reactivity $\kappa(\ell)$ such that
  \begin{equation}
  \Psi(\ell)=\exp(-D^{-1}\int_0^{\ell}\kappa(\ell')d\ell').
  \label{kapl}
  \end{equation}
  Since the probability of absorption now depends on how much time the particle spends in a neighborhood of the boundary, as specified by the local time, it follows that the stochastic process has memory. That is, absorption is non-Markovian. 

\subsection{Brownian gas on the half-line}

Writing down the SDE for an interacting Brownian gas with a partially absorbing boundary condition at $x=0$ requires that we only include particles that haven't yet been absorbed. Given a set of local time thresholds $\widehat{\bm \ell}=\{\widehat{\ell}_1,\ldots\widehat{\ell}_N\}$ and the set of stopping conditions
 \begin{equation}
\label{expj}
{\mathcal T}_j=\inf\{t>0:\ L_j(t)>\widehat{\ell}_j\},\quad \P[\widehat{\ell}_j>\ell]\equiv \Psi(\ell) ,
\end{equation}
equation (\ref{moo2}) becomes
\begin{align}
 dX_j(t)&=\bigg \{-\frac{1}{\gamma}\bigg [\partial_xV(X_j(t))+N^{-1}\sum_{k=1}^{N}\partial_x K(X_j(t)-X_k(t)){\bf 1}_{L_k(t)<\widehat{\ell}_k}\bigg ]dt\nonumber \\
 &\quad +\sqrt{2D}d{W}_j(t)+dL_j(t)\bigg \}{\bf 1}_{L_j(t)<\widehat{\ell}_j}.
\label{abs0}
 \end{align}
For simplicity, we assume that the particle absorption processes are independent and that the distributions $\Psi$ of the local time thresholds are the same for all particles. (See the discussion in section 5 for further elaboration.)
 Introduce the global density or empirical measure
 \begin{equation}
 \label{mumu}
\mu(x,\ell,\widehat{\bm \ell},t)=\frac{1}{N}\sum_{j=1}^N\mu_j(x,\ell,\widehat{\ell}_j,t)=\frac{1}{N}\sum_{j=1}^N\delta(X_j(t)-x)\delta(L_j(t)-\ell){\bf 1}_{\ell<\widehat{\ell}_j},
\end{equation}
which tracks the spatial evolution of the surviving particles given the thresholds $\widehat{\bm \ell}$. 
For a given set of local time thresholds, we derive a generalized DK equation for $\mu$ along similar lines to section 2. First, we introduce a smooth test function $f$ satisfying the constraint
$\partial_xf(0,\ell)=0$, and set
 \begin{equation}
f(X_j(t),L_j(t)){\bf 1}_{L_j(t)<\widehat{\ell}_j}=\int_{0}^{\infty}dx\int_0^{\infty}d\ell\,  \mu_j(x,\ell,\widehat{ \ell}_j,  t)f(x,\ell).
\end{equation}

Taylor expanding the composite function $f(X_j(t+dt),L_j(t+dt)){\bf 1}_{L_j(t+dt)<\widehat{\ell}_j}$ using Ito's lemma, we find that
\begin{align}
& \frac{d\left [f(X_j(t),L_j(t)){\bf 1}_{L_j(t)<\widehat{\ell}_j}\right ]}{dt}=\int_{0}^{\infty}dx\int_{0}^{\infty}d\ell\, f(x,\ell){\bf 1}_{\ell<\widehat{\ell}_j}\frac{\partial \mu_j(x,\ell,\widehat{ \ell}_j,  t)}{\partial t} \nonumber \\
 &=\int_{0}^{\infty}dx\int_{0}^{\infty}d\ell\, \mu_j(x,\ell,\widehat{\ell}_j,  t)\bigg [\sqrt{2D}\partial_x f(x,\ell)  \xi_j(t)+D\partial_{xx} f(x,\ell)+{\mathcal A}(x,\ell)\partial_xf(x,\ell) \bigg ]\nonumber \\
 &\hspace{1cm} +\int_{0}^{\infty}dx\int_{0}^{\infty}d\ell\,\delta(X_j(t)-x)\delta(L_j(t)-\ell)  D\delta(x)\partial_{\ell} \bigg [{\bm 1}_{\ell<\widehat{\ell}_j}f(x,\ell)\bigg ] , \label{beav}
\end{align}
with 
\begin{align}
\calA(x,\ell)&=-\frac{1}{\gamma } \left [\partial_xV(x)+\frac{1}{N}\sum_{k=1}^N{\bf 1}_{L_k(t)<\widehat{\ell}_k} \int_{0}^{\infty}\delta(y-X_k(t))\partial_xK(x-y)dy \right ]\nonumber \\
&=-\frac{1}{\gamma } \left [\partial_xV(x)+ \int_{0}^{\infty}\overline{\mu}(y,\widehat{\bm \ell},t)\partial_xK(x-y)dy\right ],
\end{align}
and
\begin{equation}
\overline{\mu}(x,\widehat{\bm \ell},t)=\int_0^{\infty}\mu(x,\ell,\widehat{\bm \ell},t)d\ell=\frac{1}{N}\sum_{j=1}^N\delta(X_j(t)-x){\bf 1}_{L_j(t)<\widehat{\ell}_j}.
\end{equation}
Integrating by parts the various terms gives
\begin{align}
&\int_{0}^{\infty}dx\int_{0}^{\infty}d\ell\, f(x,\ell)\frac{\partial \mu_j(x,\ell,\widehat{ \ell}_j,  t)}{\partial t} \nonumber \\
 &=\int_{0}^{\infty}dx\int_{0}^{\infty}d\ell\,\bigg [ f(x,\ell)\left (-\sqrt{2D}\partial_x\mu_j(x,\ell,\widehat{\ell}_j,t)   \xi_j(t)+D\partial_{xx} \mu_j(x,\ell,\widehat{\ell}_j,t)\right )\nonumber \\
 &\quad +\partial_x\calA(x,\ell)\mu_j(x,\ell,\widehat{\ell}_j,t)\bigg ] -\int_{0}^{\infty} \calA(0,\ell)\mu_j(0,\ell,\widehat{\ell}_j,t) d\ell \nonumber \\
&\quad -\int_{0}^{\infty} \mu_j(0,\ell,\widehat{\ell}_j,t) \left (\sqrt{2D} f(0,\ell)  \xi_j(t)+D\partial_{x} f(0,\ell)\right )d\ell +D\int_{0}^{\infty}\partial_x\mu_j(0,\ell,\widehat{\ell}_j,t) f(0,\ell)d\ell\nonumber \\
&\quad -D   \mu_j(0,0,\widehat{\ell}_j,t)  f(0,0)-D\int_{0}^{\infty}\partial_{\ell}\mu_j(0,\ell,\widehat{\ell}_j,t)f(0,\ell)d\ell+f(0,\widehat{\ell}_j)\rho_j(0,\widehat{\ell}_j,t),
\label{RHSabs}
 \end{align}
 where
 $\rho_j(0,\widehat{\ell}_j,t)=\delta(X_j(t))\delta(L_j(t)-\widehat{\ell}_j)$. The final three terms on the right-hand side arise from integrating by parts with respect to $\ell$ and using the identity
 \begin{equation}
\frac{\partial}{\partial \ell} \mu_j(x,\ell,\widehat{\ell}_j,t)= \delta(X_j(t)-x) \left [{\bf 1}_{\ell<\widehat{\ell}_j}\frac{\partial}{\partial \ell}\delta(L_j(t)-\ell)-\delta(\ell-\widehat{\ell}_j)\delta(L_j(t)-\ell)\right ].
\end{equation}

 Summing both sides of equation (\ref{RHSabs}) with respect to $j$, and transforming the noise terms along similar lines to the derivation of equation (\ref{rhoc2}) from equation (\ref{toto}), leads to the following generalized DK equation for the interacting Brownian gas:\begin{subequations}
\label{abs2int}
\begin{align} 
 \frac{\partial \mu(x,\ell,\widehat{\bm \ell},t)}{\partial t} &=-\frac{\partial J(x,\ell,\widehat{\bm \ell},t)}{\partial x},\\
 - J(0,\ell,\widehat{\bm \ell},t)&=D\frac{\partial \mu(0,\ell,\widehat{\bm \ell},t)}{\partial \ell}  +D\mu(0,0,\widehat{\bm \ell},t)  \delta(\ell)  +\frac{1}{N}\sum_{j=1}^N \rho_j(0,\widehat{\ell}_j,t)\delta(\ell-\widehat{\ell}_j) \\
 J(x,\ell,\widehat{\bm \ell},t)&=-\sqrt{\frac{2D}{N}}  \sqrt{\mu(x,\ell,\widehat{\bm \ell},t)} \eta(x,\ell,t) -D\frac{\partial \mu(x,\ell,\widehat{\bm \ell},t)}{\partial x} -H_{\mu}(x,\ell,\widehat{\bm \ell},t),
\end{align}
\end{subequations}
with
\begin{align}
H_{\mu}(x,\ell,\widehat{\bm \ell},t)&=  \gamma^{-1} \mu(x,\ell,\widehat{\bm \ell},t) \bigg (\partial_xV(x)+\int_{0}^{\infty}\overline{\mu}(y,\widehat{\bm \ell},t)\partial_xK(x-y) dy \bigg ). \end{align}
Finally, integrating equations (\ref{abs2int}) with respect to $\ell$ and simplifying the noise terms using identical arguments to the derivation of equations (\ref{rhoc1}) gives
\begin{subequations}
\label{abs2int2}
\begin{align} 
 \frac{\partial \overline{\mu}(x,\widehat{\bm \ell},t)}{\partial t} &=-\frac{\partial \overline{J}(x,\widehat{\bm \ell},t)}{\partial x},\\
  -\overline{J}(0,\widehat{\bm \ell},t)&=\overline{\nu}(\widehat{\bm \ell},t),  \\
 \overline{J}(x,\widehat{\bm \ell},t)&=-\sqrt{\frac{2D}{N}}  \sqrt{\overline{\mu}(x,\widehat{\bm \ell},t)} \eta(x,t)-D\frac{\partial \overline{\mu}(x,\widehat{\bm \ell},t)}{\partial x} -H_{\overline{\mu}}(x,\widehat{\bm \ell},t),
\end{align}
\end{subequations}
with
\begin{subequations}
\begin{align}
H_{\overline{\mu}}(x,\widehat{\bm \ell},t)&=  \gamma^{-1} \overline{\mu}(x,\widehat{\bm \ell},t) \bigg (\partial_xV(x)+\int_{0}^{\infty}\overline{\mu}(y,\widehat{\bm \ell},t)\partial_xK(x-y) dy \bigg ),\end{align}
\end{subequations}
and $\overline{\nu}({\bm \ell},t)$ an auxiliary measure defined on the boundary:
\begin{equation}
\overline{\nu}(\widehat{\bm \ell},t)=\frac{1}{N}\sum_{j=1}^N \rho_j(0,\widehat{\ell}_j,t).
\end{equation}
We see that $\overline{\nu}(\widehat{\bm \ell},t)$ represents the absorption flux through $x=0$.

In order to obtain the analog of the single particle marginal density (\ref{pmar}), we have to take expectations of equations (\ref{abs2int2}) with respect to the Gaussian noise processes and the random local time thresholds. We make use of the identities
\begin{equation}
\E[{\bf 1}_{L_j(t)<\widehat{\ell}_j}]=\Psi(L_j(t)),\quad \E[\delta(L_j(t)-\widehat{\ell}_j]=\psi(L_j(t)),
\end{equation}
and 
\begin{align}
\E\bigg [ \bigg  \langle \overline{\nu}(\widehat{\bm \ell},t)\bigg  \rangle \bigg ]&=
\E\bigg [ \bigg  \langle \frac{1}{N}\sum_{j=1}^N \delta(X_j(t))\int_0^{\infty} \delta(\ell-L_j(t))\delta(\ell-\widehat{\ell}_j)d\ell\bigg  \rangle \bigg ]\nonumber \\
&=
  \bigg  \langle \frac{1}{N}\sum_{j=1}^N  \delta(X_j(t))\psi(L_j(t))\bigg  \rangle  \nonumber \\
  &=\int_0^{\infty}\psi(\ell) \phi(0,\ell,t) d\ell ,
\end{align}
where 
\begin{equation}
\phi(x,\ell,t) =\langle \rho(x,\ell,t)\rangle =\bigg \langle \frac{1}{N}\sum_{j=1}^N \delta(x-X_j(t))\delta(\ell-L_j(t)) \bigg \rangle .
\end{equation}
Note that the particle positions $X_j(t)$ satisfy the SDE (\ref{abs0}) rather than (\ref{moo2}). Hence, $\rho(x,\ell,t)$ does not satisfy the generalized DK equation (\ref{rhoc2int}) and $\phi(x,\ell,t)$ will need to be determined another way (see below).

\begin{figure}[t!]
\centering
\includegraphics[width=16cm]{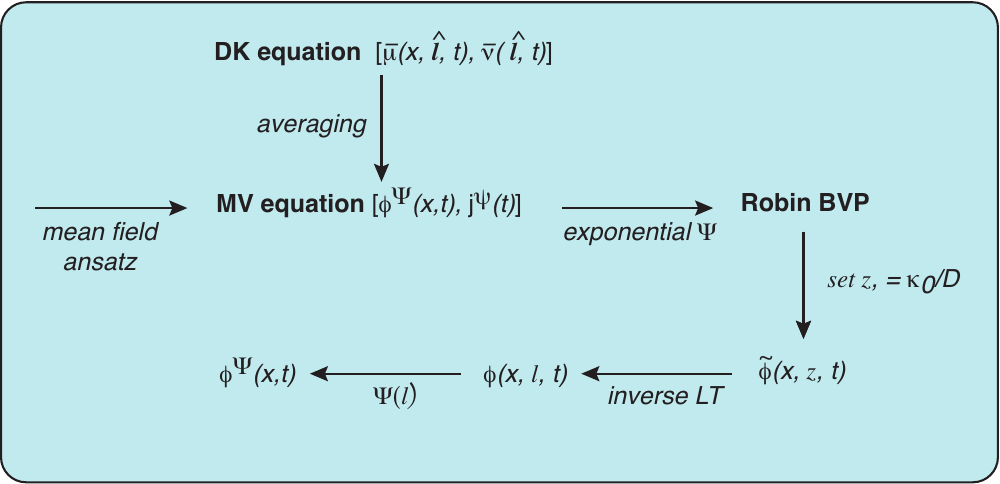} 
\caption{Hierarchy of equations obtained by combining the hydrodynamical theory of interacting Brownian gases with an encounter-based model of a partially absorbing boundary. The generalized DK equation (\ref{abs2int2}) is an SPDE for $\overline{\mu}(x,\widehat{\bm \ell},t) $ that depends on the unknown current measure $\overline{\nu}(\widehat{\bm \ell},t) $. Taking expectations with respect to the Gaussian noise processes and the local time thresholds and imposing a mean field ansatz leads to the MV equation (\ref{nob}) for the marginal density $\phi^{\Psi}(x,t)=\int_0^{\infty} \Psi(\ell) \phi(x,\ell,t)d\ell$, which couples to the absorption flux $j^{\psi}(x,t)=D\int_0^{\infty}\psi(\ell) \phi(0,\ell,t)d\ell$. In the exponential case $\Psi(\ell)=\e^{-\kappa_\ell/D}$, the MV equation reduces to a closed Robin BVP for  the marginal density, which is equivalent to the Laplace transform of $\phi(x,\ell,t)$ with respect to $\ell$. Inverting the Laplace transform then determines $\phi(x,\ell,t)$ and hence $\phi^{\Psi}(x,t)$. }
\label{fig3}
\end{figure}

The next step is assume that the mean field approximation\footnote{In the case of a partially absorbing boundary, the fraction of surviving particles is a monotonically decreasing function of time. Clearly, for a large but finite number of particles and a recurrent diffusion process, the number of remaining particles will eventually approach zero so that any mean field approximation will break down. There have been a few rigorous mathematical studies of the mean field limit and propagation of chaos for mean field games with totally absorbing boundaries \cite{Hambly17,Campi18,Caines20}.}
\begin{equation}
\E\bigg [\bigg \langle \overline{\mu}(x,\widehat{\bm \ell},t)\overline{\mu}(y,\widehat{\bm \ell},t) \bigg \rangle \bigg ]=\E\bigg [\bigg \langle \overline{\mu}(x,\widehat{\bm \ell},t)\bigg \rangle \bigg  ]\E\bigg [\bigg \langle \overline{\mu}(y,\widehat{\bm \ell},t)\bigg \rangle \bigg  ]
\end{equation} 
holds for large $N$. This then yields a MV equation for the marginal density
\begin{equation}
\phi^{\Psi}(x,t)=\E\bigg [ \bigg  \langle \overline{\mu}(x,\widehat{\bm \ell},t)\bigg  \rangle \bigg ]=\int_0^{\infty}\Psi(\ell) \phi(x,\ell,t) d\ell,
\label{o0o}
\end{equation}
which takes the form
\begin{subequations}
\label{nob}
\begin{align} 
 \frac{\partial \phi^{\Psi}(x,t)}{\partial t} &=-\frac{\partial J^{\Psi}(x,t)}{\partial x},\quad -J^{\Psi}(0,t)=j^{\psi}(t)\equiv D\int_0^{\infty} \psi(\ell)  \phi(0,\ell,t)d\ell 
 \end{align}
 with probability flux
 \begin{align}
 J^{\Psi}(x,t)&=-D\frac{\partial \phi^{\Psi}(x,t)}{\partial x} -\gamma^{-1} \phi^{\Psi}(x,t) \bigg (\partial_xV(x)+\int_{0}^{\infty}\phi^{\Psi}(y,t)\partial_xK(x-y)dy \bigg ).
\end{align}
\end{subequations}
First note that if $\Psi(\ell)=1$ for all $\ell$ then $\psi(\ell)=0$ and we recover the MV equation (\ref{MVbar2}) for a totally reflecting boundary. For almost all other choices for $\Psi$, equation (\ref{nob}) does not yield a closed PDE for $\phi^{\Psi}(y,t)$ due to the dependence of the boundary condition at $x=0$ on $\psi(\ell)$. However, as previously highlighted for single particles \cite{Grebenkov20,Grebenkov22,Bressloff22,Bressloff22a}, see also section 4.1, we do obtain a closed MV equation in the special case of an exponential distribution $\Psi(\ell) =\e^{-\kappa_0\ell/D}$, since $\psi(\ell)=\kappa_0\Psi(\ell)/D$. The boundary condition then takes the Robin form
\begin{equation}
J^{\Psi}(0,t)=-\kappa_0 \phi^{\Psi}(0,t),
\end{equation}
where $\kappa_0$ is the effective reactivity. It immediately follows by analogy with the single particle case, that if we can solve the nonlinear Robin BVP then we can interpret the solution as the Laplace transform $\widetilde{\phi}(x,z=\kappa_0/D,t)$ of $\phi(x,\ell,t)$ with respect to $\ell$. Inverting this Laplace transform then determines $\phi(x,\ell,t)$ and hence $\phi^{\Psi}(x,t)$ for general $\Psi$ according to equation (\ref{o0o}). The relationships between the various equations obtained by combining the hydrodynamics of an interacting Brownian gas and an encounter-based model of partially absorbing boundaries are summarized in Fig. \ref{fig3}.

Finally, as in the case of reflecting boundaries, it is possible to extend our results to an interacting
Brownian gas confined on the interval $[-R,R]$ with a partially absorbing boundary at each end. The details of the absorption process will depend on whether or not we have separate thresholding conditions at the two ends (see the discussion in section 5). Here, we consider the simpler case in which absorption of the $j$th particle occurs as soon as the joint local time $L_j(t)$ given by equation (\ref{local2}) exceeds the random threshold $\widehat{\ell}_j$, irrespective of which end this occurs. The corresponding MV equation is
\begin{subequations}
\label{nob2}
\begin{align} 
 \frac{\partial \phi^{\Psi}(x,t)}{\partial t} &=-\frac{\partial J^{\Psi}(x,t)}{\partial x},\quad J^{\Psi}(\pm R,t)=\pm D\int_0^{\infty} \psi(\ell)  \phi(\pm R,\ell,t)d\ell ,
 \end{align}
 with probability flux
 \begin{align}
 J^{\Psi}(x,t)&=-D\frac{\partial \phi^{\Psi}(x,t)}{\partial x} -\gamma^{-1} \phi^{\Psi}(x,t) \bigg (\partial_xV(x)+\int_{-R}^{R}\phi^{\Psi}(y,t)\partial_xK(x-y)dy \bigg ).
\end{align}
\end{subequations}

\subsection{Weak absorption limit}

As a simple example, consider the half-line with the potential $V(x)=\nu x^2/2$ and take the interactions to be given by the Curie-Weiss potential  $K(x)=\lambda x^2/2$, $\lambda >0$. We also choose the exponential distribution $\Psi(x)=\e^{-\kappa_0\ell/D}$ with $\kappa_0 \ll D/L$, so that absorption is much slower than diffusion. The probability flux becomes (after dropping the superscript $\Psi$)
\begin{align}
 J(x,t)&=-D\frac{\partial \phi(x,t)}{\partial x} -\frac{1}{\gamma}V'(x)\phi(x,t) -\frac{\lambda}{\gamma}\phi(x,t) \int_{0}^{\infty}\phi(y,t)(x-y)dy \nonumber \\
 &= -D\frac{\partial \phi(x,t)}{\partial x} -\frac{1}{\gamma}\phi(x,t) [V'(x)+\lambda \Gamma(t)x-\lambda m(t)]
\end{align}
with 
\begin{equation}
\label{mt}
m(t)=\int_{0}^{\infty}y\phi(y,t)dy.
\end{equation} 
We thus obtain the non-autonomous FP equation 
\begin{subequations}
\label{Robin2}
\begin{align} 
 \frac{\partial \phi(x,t)}{\partial t} &=-\frac{\partial J(x,t)}{\partial x}=D\frac{\partial^2 \phi(x,t)}{\partial x^2}+\frac{1}{\gamma} \frac{\partial [A(x,t) \phi(x,t)]}{\partial x} ,\ x\in (0,\infty),\\
 J(0,t)&= - \epsilon \kappa_0\phi(0,t) ,\end{align}
 \end{subequations}
 with
 \begin{equation}
 A(x,t)=\lambda \,m(t)-V'(x)-\lambda x \Gamma(t).
 \end{equation}
We have rescaled the reactivity by the small positive parameter $\epsilon$ in order to reflect the relative slow rate of absorption. We wish to calculate the loss function
\begin{equation}
\Lambda(t)\equiv   \int_{0}^{\infty}  \phi(x,t)  dx,
\label{GPsi}
\end{equation}
which is the expected fraction of particles that have not been absorbed up to time $t$. (It is the analog of the survival probability for a single particle.) It follows from equations (\ref{Robin2}a) that
\begin{equation}
\frac{d\Lambda(t)}{dt}=-\epsilon \kappa_0 \phi(0,t) .
\label{GPsi2}
\end{equation}

\begin{figure}[t!]
\centering
\includegraphics[width=16cm]{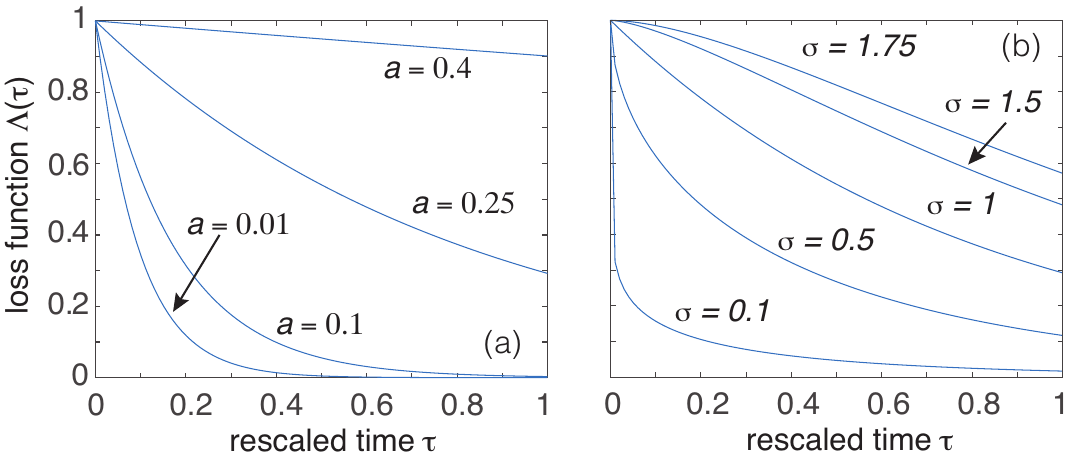} 
\caption{Loss function in the weak absorption limit for an interacting Brownian gas on the half-line. (a) Plot of the exponential loss function $\Lambda(\tau)=\e^{-\tau/Z(a)}$ for various values of the first moment $a$. (b) Plot of loss function $\Lambda(\tau)=\Psi_{\rm gam}( \tau/Z(a))$ for $\psi_{\rm gam}(\ell)=-\Psi_{\rm gam}'(\ell)$ given by the gamma distribution (\ref{psigam}) and various values of the parameter $\sigma$. The first moment is $a=0.25$. Other parameters are $D=\kappa_0=1$.}
\label{fig4}
\end{figure}

In order to solve equation (\ref{Robin2}), we exploit the fact that when $\epsilon=0$ there exists a unique stationary state $\phi=\phi_a(x)$, $m=m(a)=a$ and $\Lambda=1$, see section 3a. Using a quasi-steady-state approximation, we introduce the slow time-scale $\tau=\epsilon t$ and set
\begin{equation}
\phi(x,t)=\Lambda(\tau)\phi_a(x)+\epsilon u(x,\tau),
\end{equation}
with $\int_{0}^{\infty}u(x,\tau)dx=0$, $\Lambda(0)=1$ and $u(x,0)=0$. In particular, we assume that the systems starts in the stationary state of the reflecting BVP.
Substituting into equation (\ref{GPsi2}) shows that to leading order
\begin{align} 
  \frac{d\Lambda(\tau)}{d\tau}&= -   \kappa_0 \phi_a(0)\Lambda(\tau) ,
  \end{align}
  so that
  \begin{equation}
  \phi(x,t)\sim \e^{-\epsilon  \kappa_0 \phi_a(0) t}\phi_a(x).
  \end{equation}
  Equations (\ref{ssbar}) and (\ref{Za}) imply that
  \begin{align}
\phi_a(0)=\frac{1}{Z(a)},\quad Z(a)
=\sqrt{\frac{\pi}{2\beta[ \nu +\lambda]}}\e^{\beta a^2\lambda^2/2[\nu +\lambda]}\mbox{erfc}(-a\lambda \sqrt{\beta/2[\nu+\lambda]}),
\label{pa}
\end{align}
with $a$ determined from equation (\ref{acon}). Example plots of the loss function $\Lambda(\tau)=\e^{- \tau/Z(a)}$ are shown in Fig. \ref{fig4}(a) for various first moments $a$. We absorb the constant $\kappa_0$ into the slow time $\tau$.
  Finally, given the relationship between the solution to the Robin BVP and the Laplace transformed propagator, we can set
   \begin{equation}
\phi(x,\ell,t)\sim \delta(\ell-\epsilon D \phi_a(0) t)\phi_a(x),
\end{equation} 
and for a general local time threshold distribution
\begin{equation}
\phi^{\Psi}(x,t)\sim \Psi(\epsilon D \phi_a(0)t)\phi_a(x).
\end{equation}

One example of a non-exponential threshold distribution is the gamma distribution
\begin{equation}
\label{psigam}
\psi_{\rm gam}(\ell)=\frac{r(r \ell)^{\sigma-1}\e^{-r  \ell}}{\Gamma(\sigma)},\quad \Psi_{\rm gam}(\ell)=\frac{\Gamma(\sigma,r \ell)}{\Gamma(\sigma)} ,\sigma >0,\quad r=\kappa_0/D,
\end{equation}
where $\Gamma(\sigma)$ is the gamma function and $\Gamma(\sigma,z)$ is the upper incomplete gamma function:
\begin{equation}
\Gamma(\sigma)=\int_0^{\infty}\e^{-t}t^{\sigma-1}dt,\quad \Gamma(\sigma,z)=\int_z^{\infty}\e^{-t}t^{\sigma-1}dt,\ \sigma>0.
\end{equation}
The parameter $r$ determines the effective absorption rate so that the boundary $x=0$ is non-absorbing in the limit $r \rightarrow 0$ and totally absorbing in the limit $r\rightarrow \infty$. If $\sigma=1$ then $\psi_{\rm gam}$ reduces to the exponential distribution $\psi_{\rm gam}(\ell)|_{\sigma =1}=r \e^{-r \ell}$. The parameter $\sigma$ thus characterizes the deviation of $\psi_{\rm gam}(\ell)$ from the exponential case. If $\sigma <1$ ($\sigma>1$) then $\Psi(\ell)$ decreases more rapidly (slowly) as a function of the local time $\ell$, that is, the boundary is more (less) absorbing. Example plots of the corresponding loss function $\Lambda(\tau)=\Psi_{\rm gam} (\tau/Z(a))$ are shown in Fig. \ref{fig4}(b). We fix $D=1$ and $a=0.25$, and consider various values of the gamma distribution parameter $\sigma$. It can be seen that as the boundary becomes more absorbing (decreasing $\sigma$), the loss function decays more rapidly. Moreover, $\Lambda(\tau)$ is a convex down (up) function of $\tau$ for $\sigma >1$ ($\sigma <1$).

\section{Summary and extensions}
In this paper we considered the problem of an interacting Brownian gas in the semi-infinite and finite intervals. In order to handle the boundary conditions, we introduced a global density that keeps track of both the positions and boundary local times of all of the surviving particles (in the case of partial absorption). We derived generalized DK equations for the global density, see equations (\ref{rhoc2int}) and (\ref{abs2int}), which are exact SPDEs that prescribe how to incorporate the effects of spatiotemporal noise at the population level. Although the resulting DK equations are exact, their solutions are highly singular. Therefore, we used a mean field ansatz to reduce the DK equations to nonlinear MV equations in the thermodynamic limit, see equations (\ref{MVbar2}) and (\ref{nob}). The rigorous mathematical proof that the mean field limit exists via propagation of chaos has been carried out in the case of reflecting boundaries \cite{Sznitman84,Coghi22} but not, as far as we are aware, for partially absorbing boundaries. The latter also has the additional complication that there is a constant loss of particles due to absorption, so that for a large but finite number of particles and recurrent diffusion, any mean field approximation will eventually break down. Independently of these particular issues, the boundary value problems for the generalized DK equation and the MV equations are of intrinsic interest. The former provides a starting point for developing various approximation schemes for large but finite populations, whereas the latter is an example of a nonlinear, nonlocal PDE with rich mathematical structure.

One natural direction for future work is to consider higher-dimensional versions of interacting Brownian gases in bounded domains. Here we briefly discuss some particular extensions of the encounter-based approach.
\medskip

\begin{figure}[b!]
  \centering   \includegraphics[width=16cm]{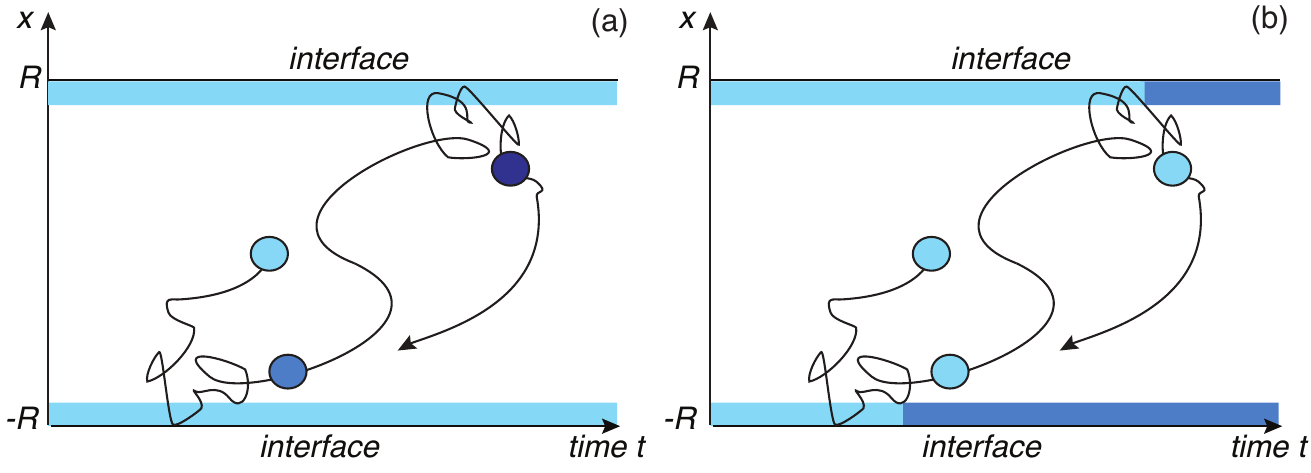}
  \caption{Schematic diagram indicating two different absorption scenarios for a Brownian particle diffusing in the interval $[-R,R]$. For the sake of illustration, the local times are taken to be the amount of time spent in a small boundary layer around $x=\pm R$. (a) The particle has an internal state $S(t)$ that increases strictly monotonically with the amount of time $L(t)$ spent in contact with either boundary. Absorption occurs when the internal state, and hence $L(t)$, crosses a threshold. (b) Each boundary has its own internal state denoted by $S^{\pm}(t)$, which is a strictly monotonic function of the amount of time the boundary is in contact with the particle, which is specified by the local time $L^{\pm}(t)$. Absorption occurs as soon as one of the internal states crosses its corresponding threshold. In both cases (a) and (b), the value of the internal state is represented by the color of the shaded regions.}
  \label{fig5}
  \end{figure}

\noindent {\bf (i) Independently absorbing boundaries.} In section 4 we assumed that each particle is independently absorbed according to the stopping conditions (\ref{expj}), with all particles having the same local time threshold distribution $\Psi$. A simple generalization would be to take the distributions to be $j$-dependent. A related issue is that, in the case of the finite interval $[-R,R]$, we did not distinguish between absorption events at the two ends $x=\pm R$. That is $L_j(t)$ was taken to be the sum of the local times accrued at both ends, see equation (\ref{local2}). An alternative model would treat the absorption processes at $x=\pm R$ to be independent. At the single particle level, this would mean introducing the pair of local times
\begin{equation}
L^{+}(t)=\lim_{\epsilon\rightarrow 0^+}\frac{D}{\epsilon} \int_0^tI_{(R-\epsilon,R)}(X(s))ds,\quad L^-(t)=\int_0^tI_{(-R,-R+\epsilon)}(X(s))ds 
\end{equation}
and the modified stopping condition
\begin{equation}
\calT=\min\{\calT^-,\calT^+\},\quad \calT^{\pm} = \inf\{t>0:\ L^{\pm}(t)>\widehat{\ell}^{\pm}\}, \quad \P[\widehat{\ell}^{\pm}>\ell]\equiv \Psi{\pm}(\ell) .
\end{equation}
The difference between the two scenarios also has a possible physical interpretation as illustrated in Fig. \ref{fig5}. In particular, recall that if $\Psi$ is non-exponential, then the absorption process is non-Markovian, that is, some memory trace of previous particle-boundary encounters is maintained. Treating each particle as independently absorbed suggests that the memory traces are associated with internal states of the particles, see Fig. \ref{fig5}(a). However, another possibility is that the individual boundaries maintain the memory traces, see Fig. \ref{fig5}(b) so that the absorption process at the two ends can be separated. However, the latter significantly complicates the analysis of the multi-particle Brownian gas, since the probability that any one particle is absorbed will depend on previous interactions between the boundary and all other particles. 
\medskip

\begin{figure}[t!]
\centering
   \includegraphics[width=16cm]{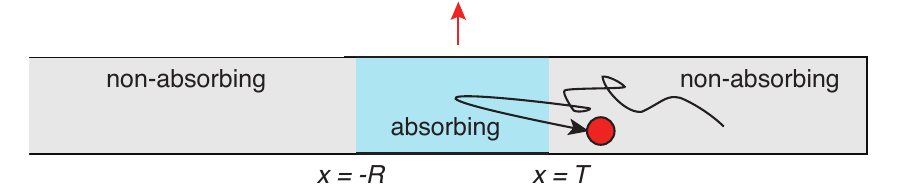}
  \caption{One-dimensional diffusion with a partially absorbing trap in the interval $[-R,R]$.}
  \label{fig6}
  \end{figure}
  
\noindent {\bf (ii) Single particle diffusion with a partially absorbing trap.} The encounter-based approach to single particle absorption has also been developed within the context of heterogeneous media, where one or more subregions of a domain act as partially absorbing traps \cite{Bressloff22,Bressloff22a}. This is illustrated in Fig. \ref{fig6} for an absorbing trap in the interval $[-R,R]$. A Brownian particle can freely enter and exit the trap but is only absorbed within the trap when its occupation time exceeds some random threshold.
The occupation time is a Brownian functional defined according to \cite{Majumdar05} 
\begin{equation}
\label{occ}
A(t)=\int_{0}^t{\bf 1}_{(-R,R)}(X(\tau))d\tau .
\end{equation}
$A(t)$ specifies the amount of time the particle spends within $[-R,R]$ over the time interval $[0,t]$. Denoting the generalized propagator by $P(x,a,t)$ and $Q(x,a,t)$ for $x\notin (-R,R)$ and $x\in (-R,R)$, respectively, we have the BVP \cite{Bressloff22a}
\begin{subequations}
\begin{align}
\label{Pocca}
&\frac{\partial P(x,a,t )}{\partial t}=D\frac{\partial^2 P(x,a,t )}{\partial x^2}, \ |x|\geq R,\\
 &\frac{\partial Q(x,a,t)}{\partial t}=D\frac{\partial^2 Q(x,a,t)}{\partial x^2} -\left (\frac{\partial Q}{\partial a}(x,a,t) +\delta(a)Q(x,0,t) \right ),\ -R<x<R.
\label{Poccb}
\end{align}
These are supplemented by matching conditions at the interfaces $x=\pm R$,
	\begin{equation}
	\label{Poccd}
 P(\pm R,a,t)=Q(\pm R,a,t),\quad  \left .\frac{\partial P(x,a,t)}{\partial x}\right |_{x=\pm R}=\left .\frac{\partial Q(x,a,t)}{\partial x}\right |_{x=\pm R}.
	\end{equation}
	\end{subequations}
Finally, the stopping time condition is
\begin{equation}
\label{TA}
{\mathcal T}=\inf\{t>0:\ A(t) >\widehat{a}\},
\end{equation}
where $\widehat{a}$ is a random variable with probability distribution $\Psi(a)=\P[\widehat{a}>a]$. The marginal probability density for particle position $X(t) $ is then \cite{Bressloff22a}
\begin{subequations}
\begin{eqnarray}
\label{peep}
p(x,t)&=&\int_0^{\infty}\Psi(a) P(x,a,t)da,\ |x|\geq R,\\
q(x,t)&=&\int_0^{\infty}\Psi(a) Q(x,a,t)da,\ -R \leq x \leq R.
\end{eqnarray}
\end{subequations}

\noindent {\bf (iii) Interacting Brownian gas with a partially absorbing trap.}  Incorporating a partially absorbing trap into a model of a one-dimensional interacting Brownian gas proceeds along analogous lines to the case of a partially absorbing boundary on the half-line, with $L_j(t)$ replaced by $A_j(t)$. Here we sketch the basic steps, leaving the details to future work. First, the SDE (\ref{abs0}) is replaced by 
\begin{align}
 dX_j(t)&=\bigg \{-\frac{1}{\gamma}\bigg [\partial_xV(X_j(t))+N^{-1}\sum_{k=1}^{N}\partial_x K(X_j(t)-X_k(t)){\bf 1}_{A_k(t)<\overline{a}_k}\bigg ]dt\nonumber \\
 &\quad +\sqrt{2D}d{W}_j(t) \bigg \}{\bf 1}_{A_j(t)<\overline{a}_j}.
\label{trap}
 \end{align}
 Second, the global density is defined according to
 \begin{equation}
 \label{mutrap}
\mu(x,a,\widehat{\bm a},t)=\frac{1}{N}\sum_{j=1}^N \mu_j(x,a,\widehat{ a}_j,  t)= \frac{1}{N}\sum_{j=1}^N\delta(X_j(t)-x)\delta(A_j(t)-a){\bf 1}_{a<\widehat{a}_j}.
\end{equation}
Introduce a smooth test function $f(x,a)$ for $x\in \R$ and $a\in [0,\infty)$, and set
 \begin{equation}
f(X_j(t),A_j(t)){\bf 1}_{A_j(t)<\widehat{a}_j}=\int_{-\infty}^{\infty}dx\int_0^{\infty}da\,  \mu_j(x,a,\widehat{ a}_j,  t)f(x,a).
\end{equation}
Applying Ito's lemma to this equation, integrating by parts, summing over $j$, and transforming the noise terms leads to the following
generalized DK equation for $\mu$:
\begin{subequations}
\label{trapint0}
\begin{align} 
 \frac{\partial \mu(x,a,\widehat{\bm a},t)}{\partial t} &=-\frac{\partial J(x,a,\widehat{\bm a},t)}{\partial x}- {\mathcal J}(x,a,\widehat{\bm a},t),\\
 J(x,a,\widehat{\bm a},t)&=-\sqrt{\frac{2D}{N}}  \sqrt{\mu(x,a,\widehat{\bm a},t)} \eta(x,a,t) -D\frac{\partial \mu(x,a,\widehat{\bm a},t)}{\partial x} -H_{\mu}(x,a,\widehat{\bm a},t),
\end{align}
\end{subequations}
with
\begin{align}
H_{\mu}(x,a,\widehat{\bm a},t)&=  \gamma^{-1} \mu(x,a,\widehat{\bm a},t) \bigg (\partial_xV(x)+\int_{-\infty}^{\infty}\overline{\mu}(y,\widehat{\bm a},t)\partial_xK(x-y) dy \bigg ), \end{align}
$\overline{\mu}(x,\widehat{\bm a},t)=\int_0^{\infty}\mu(x,a,\widehat{\bm a},t) da$, and
\begin{eqnarray}
&&{\mathcal J}(x,a,\widehat{\bm a},t)\\
&&=\int_{-R}^R\delta(x-y)\left [\frac{\partial \mu(y,a,\widehat{\bm a},t)}{\partial a}+ \mu(y,0,\widehat{\bm a},t)\delta(a)+\frac{1}{N}\sum_{j=1}^N \rho_j(y,\widehat{a}_j,t)\delta(a-\widehat{a}_j)\right ]dy.\nonumber 
\end{eqnarray}
Integrating with respect to $a$ and simplifying the noise terms yields a corresponding DK equation for $\overline{\mu}$
\begin{subequations}
\label{trapint}
\begin{align} 
 \frac{\partial \overline{\mu}(x,\widehat{\bm a},t)}{\partial t} &=-\frac{\partial \overline{J}(x,\widehat{\bm a},t)}{\partial x}-\overline{\nu}(x,\widehat{\bm a},t),\\
   \overline{J}(x,\widehat{\bm a},t)&=-\sqrt{\frac{2D}{N}}  \sqrt{\overline{\mu}(x,\widehat{\bm a},t)} \eta(x,t)-D\frac{\partial \overline{\mu}(x,\widehat{\bm a},t)}{\partial x} -H_{\overline{\mu}}(x,\widehat{\bm a},t),
\end{align}
\end{subequations}
with
\begin{subequations}
\begin{align}
H_{\overline{\mu}}(x,\widehat{\bm a},t)&=  \gamma^{-1} \overline{\mu}(x,\widehat{\bm a},t) \bigg (\partial_xV(x)+\int_{-\infty}^{\infty}\overline{\mu}(y,\widehat{\bm a},t)\partial_xK(x-y) dy \bigg ),\end{align}
\end{subequations}
and 
\begin{equation}
\overline{\nu}(x,\widehat{\bm a},t)=\frac{1}{N}\int_{-R}^R\delta(x-y) \sum_{j=1}^N \rho_j(y,\widehat{a}_j,t)dy.
\end{equation}
We see that $\overline{\nu}(x,\widehat{\bm a},t)$, $x\in (-R,R)$, represents the absorption flux within the trap region.

The final step is to take expectations with respect to the Gaussian noise and occupation time thresholds. Under a mean field ansatz, we obtain an MV equation for the marginal density
\begin{equation}
\phi^{\Psi}(x,t)=\E\bigg [ \bigg  \langle \overline{\mu}(x,\widehat{\bm a},t)\bigg  \rangle \bigg ]=\int_0^{\infty}\Psi(a) \phi(x,a,t) da,
\end{equation}
which takes the form
\begin{subequations}
\label{nobby}
\begin{align} 
 \frac{\partial \phi^{\Psi}(x,t)}{\partial t} &=-\frac{\partial J^{\Psi}(x,t)}{\partial x}-\int_{-R}^R \delta(x-y) j^{\psi}(y,t),
  \end{align}
 with probability flux
 \begin{align}
 J^{\Psi}(x,t)&=-D\frac{\partial \phi^{\Psi}(x,t)}{\partial x} -\gamma^{-1} \phi^{\Psi}(x,t) \bigg (\partial_xV(x)+\int_{-\infty}^{\infty}\phi^{\Psi}(y,t)\partial_xK(x-y)dy \bigg ),
\end{align}
and absorption flux
\begin{equation}
j^{\psi}(y,t)=\int_0^{\infty}\psi(a)\phi(y,a,t)da, \quad y\in (-R,R).
\end{equation}
\end{subequations}
As in the case of partially absorbing boundaries, equation (\ref{nobby}) reduces to a closed equation for $\phi^{\Psi}(x,t)$ for the exponential distribution $\Psi(a)=\e^{-\kappa_0 a}$ since $j^{\psi}(y,t)\rightarrow \kappa_0 \phi^{\Psi}(y,t)$. That is, there is a constant rate of absorption $\kappa_0$ within the trap. Assuming that a solution to the resulting nonlinear BVP can be found, then it is equivalent to the Laplace transform $\widetilde{\phi}(x,z,t)$ of $\phi(x,a,t)$ with $z=\kappa_0$. Inverting this Laplace transform then determines $\phi(x,a,t)$ and hence $p^{\Psi}(x,t)$ for a general distribution $\Psi$.

\medskip

\setcounter{equation}{0}
\renewcommand{\theequation}{A.\arabic{equation}}
\section*{Appendix A: SPDE for the equal-time correlation function}

In this appendix we derive an SPDE for the product
$C(x,y,\ell,\ell',t) \equiv \rho(x,\ell,t)\rho(y,\ell',t)$. 
First, consider the decomposition
\begin{equation}
N^2 C(x,y,\ell,\ell',t)=\sum_{i=1}^{N}\left [\sum_{j\neq i}C_{ij}(x,y,\ell,\ell',t)+C_{ii}(x,y,\ell,\ell')\right ],
\end{equation}
\begin{equation}
C_{ij}(x,y,\ell,\ell',t)\equiv \rho_i(x,\ell,t)\rho_j(y,\ell',t),\ j\neq i, \quad C_{ii}(x,y,\ell,\ell',t)=\delta(x-y)\delta(\ell-\ell')\rho_i(x,\ell,t).
\end{equation}
In order to simplify the notation, we set ${\bf z}=(x,y,\ell,\ell')$ and 
\begin{equation}
\iint d{\bf z}=\int_0^{\infty}dx\int_0^{\infty}dy \int_0^{\infty}d\ell\int_0^{\infty}d\ell' .
\end{equation}
Introduce the test function $f(x,y,\ell,\ell')$ such that $\partial_xf(0,y,\ell,\ell')=0$ for all $y\geq 0$ and $\partial_yf(x,0,\ell,\ell')=0$ for all $x\geq 0$. Applying Ito's lemma to $f(t)=f(X_i(t),X_j(t),L_i(t),L_j(t))$ and using
\begin{equation}
f(X_i(t),X_j(t),L_i(t),L_j(t))=\iint d{\bf z} \, f(\z)C_{ij}(\z,t),
\end{equation}
yields
\begin{align}
\label{Adfl}
 \iint d\z \, f(\z)\frac{\partial C_{ij}(\z,t)}{\partial t} &=\iint d\z \, C_{ij}(\z,t)\bigg [\sqrt{2D}\partial_x f(\z)  \xi_i(t)+D\partial_{xx} f(\z)+D\partial_{\ell}f(\z)\delta(x)\bigg ]\nonumber \\
 &\quad +\iint d\z \, C_{ij}(\z,t)\bigg [\sqrt{2D}\partial_y f(\z)  \xi_j(t)+D\partial_{yy} f(\z)+D\partial_{\ell'}f(\z)\delta(y)\bigg ]\nonumber \\
 &\quad +2D \delta_{i,j} \iint d\z \, C_{ii}(\z,t) \partial_{xy} f(\z)
\end{align}
The last term follows from $dW_i(t)dW_j(t)=\delta_{i,j}dt$. Performing integration by parts, we find that
\begin{equation}
 \iint  d\z \, f(\z)\frac{\partial C_{ij}(\z,t)}{\partial t} ={\mathcal I}_1(t)+{\mathcal I}_2(t)
\end{equation}
with
\begin{align*}
{\mathcal I}_1(t) &=\iint d\z \, f(\z)\bigg [-\sqrt{2D}\partial_x C_{ij}(\z,t)  \xi_i(t)+D\partial_{xx}C_{ij}(\z,t)-D\partial_{\ell}C_{ij}(\z,t)\delta(x)\bigg ] \nonumber \\
 &\quad +\iint d\z \, f(\z)\bigg [-\sqrt{2D}\partial_y C_{ij}(\z,t)  \xi_j(t)+D\partial_{yy} C_{ij}(\z,t)-D\partial_{\ell'}C_{ij}(\z,t)\delta(y)\bigg ]\nonumber \\
 &\quad +2D \delta_{i,j} \iint d\z \, f(\z) \partial_{xy} C_{ii}(\z,t) -\iint d\z\, \delta(x) C_{ij}(\z,t)\left (\sqrt{2D} f(\z)  \xi_i(t)+D\partial_{x} f(\z)\right )
 \end{align*}
 and
 \begin{align*}
 {\mathcal I}_1(t)&=  -\iint d\z\, \delta(y) C_{ij}(\z,t)\left (\sqrt{2D} f(\z)  \xi_j(t)+D\partial_{y} f(\z)\right )\nonumber \\
&\quad +D\iint d\z\, \delta(x) \partial_xC_{ij}(\z,t)f(\z)+D\iint d\z\, \delta(y) \partial_yC_{ij}(\z,t)f(\z)\nonumber \\
&\quad -D\iint d\z\,  \delta(x)\delta(\ell) C_{ij}(\z,t) f(\z) -D\iint d\z\,  \delta(y)\delta(\ell')C_{ij}(\z,t) f(\z)  \nonumber \\
&\quad  +2D\delta_{i,j}\iint d\z\, \delta(x) \partial_yC_{ij}(\z,t)f(\z) +2D\delta_{i,j} \iint d\z\, \delta(y) \partial_xC_{ij}(\z,t) f(\z) .
\end{align*}
Imposing the boundary conditions $\partial_xf(0,y,\ell,\ell')=0$ and $\partial_yf(x,0,\ell,\ell')=0$ and exploiting the arbitrariness of $f$ otherwise, we obtain the following SPDE for $C_{ij}$:
\begin{subequations}
\label{Cijeq}
\begin{align}
\frac{\partial C_{ij}(x,y,\ell,\ell',t)}{\partial t}&=-\sqrt{2D}\frac{\partial C_{ij}(x,y,\ell,\ell',t)}{\partial x}   \xi_i(t)-\sqrt{2D}\frac{\partial C_{ij}(x,y,\ell,\ell',t)}{\partial y}   \xi_j(t)\nonumber \\
&\quad +D\frac{\partial^2C_{ij}(x,y,\ell,\ell',t)}{\partial x^2}+D\frac{\partial^2C_{ij}(x,y,\ell,\ell',t)}{\partial y^2}+2D\delta_{i,j}\frac{\partial^2C_{ij}(x,y,\ell,\ell',t)}{\partial x\partial y}\nonumber \\
&\quad  +\delta(x){\mathcal J}^{(1)}_{ij}(y,\ell,\ell',t)+\delta(y){\mathcal J}^{(2)}_{ij}(x,\ell,\ell',t),
\end{align}
with
\begin{align}
{\mathcal J}_{ij}^{(1)}(y,\ell,\ell',t)&\equiv D\frac{\partial C_{ij}(0,y,\ell,\ell',t)}{\partial x}-D\frac{\partial C_{ij}(0,y,\ell,\ell',t)}{\partial \ell} -\sqrt{2D} C_{ij}(0,y,\ell,\ell',t) \xi_i(t)\nonumber \\
&\quad -DC_{ij}(0,y,0,\ell',t)  \delta(\ell)+2D\delta_{i,j} \frac{\partial C_{ij}(0,y,\ell,\ell',t)}{\partial y} , \\
{\mathcal J}_{ij}^{(2)}(x,\ell,\ell',t)&\equiv D\frac{\partial C_{ij}(x,0,\ell,\ell',t)}{\partial y}-D\frac{\partial C_{ij}(x,0,\ell,\ell',t)}{\partial \ell'} -\sqrt{2D} C_{ij}(x,0,\ell,\ell',t) \xi_j(t)\nonumber \\
&\quad -DC_{ij}(x,0,\ell,0,t)  \delta(\ell')  +2D\delta_{i,j} \frac{\partial C_{ij}(x,0,\ell,\ell',t)}{\partial x}  .
\end{align}
\end{subequations}
Finally, summing equations (\ref{Cijeq}) with respect to $i,j$ and dividing through by $N^2$, we have
\begin{subequations}
\label{Ceq}
\begin{align}
\frac{\partial C(x,y,\ell,\ell',t)}{\partial t}&=-\frac{\sqrt{2D}}{N^2}\sum_{i,j=1}^{N}\left [\frac{\partial C_{ij}(x,y,\ell,\ell',t)}{\partial x}   \xi_i(t)+ \frac{\partial C_{ij}(x,y,\ell,\ell',t)}{\partial y}   \xi_j(t)\right ]\nonumber \\
&\quad +D\frac{\partial^2C(x,y,\ell,\ell',t)}{\partial x^2}+D\frac{\partial^2C(x,y,\ell,\ell',t)}{\partial y^2}\nonumber \\
&\quad +\frac{2D}{N}\delta(\ell-\ell')\frac{\partial^2}{\partial x\partial y}\delta(x-y)\rho(x,\ell,t),
\end{align}
together with the boundary conditions
\begin{align}
 D\frac{\partial C(0,y,\ell,\ell',t)}{\partial x}&=D\frac{\partial C(0,y,\ell,\ell',t)}{\partial \ell} +\frac{\sqrt{2D}}{N^2} \sum_{i,j=1}^{N}C_{ij}(0,y,\ell,\ell',t) \xi_i(t)\nonumber \\
&\quad +DC(0,y,0,\ell',t)  \delta(\ell),\ y>0, \\
D\frac{\partial C(x,0,\ell,\ell',t)}{\partial y}&=D\frac{\partial C(x,0,\ell,\ell',t)}{\partial \ell'} +\frac{\sqrt{2D}}{N^2} \sum_{i,j=1}^{N}C_{ij}(x,0,\ell,\ell',t) \xi_j(t)  \nonumber \\
&\quad +D C(x,0,\ell,0,t)  \delta(\ell') ,\quad x>0  .
\end{align}
\end{subequations}
Averaging these equations with respect to the spatiotemporal white noise leads to the deterministic PDE (\ref{detCeq}).

\end{document}